\documentclass[useAMS,usenatbib]{mn2e}

\usepackage{caption}
\usepackage{siunitx}
\usepackage{multirow}
\usepackage[fleqn]{amsmath}
\usepackage{lipsum}
\usepackage{natbib}
\usepackage{amssymb}
\usepackage{graphicx}
\usepackage{adjustbox}
\newsavebox{\measurebox}

\title[Evolution of star cluster systems in galaxies]{Evolution of star cluster systems in isolated galaxies: \\first results from direct $N$-body simulations}

\author[Rossi, Bekki \& Hurley]{L. J. Rossi$^{1}$$^{,}$$^{2}$\thanks{E-mail: lucarossi@swin.edu.au}, K. Bekki$^{2}$, J. R. Hurley$^{1}$\\
$^{1}$ Centre for Astrophysics and Supercomputing, Swinburne University of Technology, Hawthorn, VIC 3122, Australia\\
$^{2}$ ICRAR, M468, The University of Western Australia, 35 Stirling Hwy, Crawley, WA 6009, Australia}

\begin{document}

\date{Accepted ----. Received ----; in original form ----}

\pagerange{\pageref{firstpage}--\pageref{lastpage}} \pubyear{2015}

\maketitle

\label{firstpage}

\begin{abstract}
The evolution of star clusters is largely affected by the tidal field generated by the host galaxy. It is thus in principle expected that under the assumption of an ``universal''  initial cluster mass function the properties of the evolved present-day mass function of star cluster systems should show a dependency on the properties of the galactic environment in which they evolve. To explore this expectation a sophisticated model of the tidal field is required in order to study the evolution of star cluster systems in realistic galaxies. Along these lines, in the present work we first describe a method developed for coupling $N$-body simulations of galaxies and star clusters. We then generate a database of galaxy models along the Hubble sequence and calibrate evolutionary equations to the results of direct $N$-body simulations of star clusters in order to predict the clusters' mass evolution  as function of the galactic environment. We finally apply our methods to explore the properties of evolved ``universal'' initial cluster mass functions and any dependence on the host galaxy morphology and mass distribution. The preliminary results show that an initial power-law distribution of the masses ``universally'' evolves into a log-normal  distribution, with the properties correlated with the stellar mass and stellar mass density density of the host galaxy.
\end{abstract}


\section{Introduction}
\label{sec:introduction}
Star clusters (SCs) and galaxies represent an example of dynamical stellar systems in which formation and evolution are strongly interconnected. The physical mechanisms leading to the formation of SCs in galaxies have been the subject of numerous studies \citep[][amongst others]{Peebles68,Peebles84,Kumai93,Elmegreen96,Abel98,Ashman01,Weil01,Vandenbergh01,
Bekki02,Beasley02,Bekki05,Kravtsov05,Bekki07,Griffen10,Kruijssen11,Kruijssen12,Katz14,Kruijssen15}. The results of these works suggest that star clusters form in the cores of giant molecular clouds, both at high redshifts (old globular clusters) and in the local Universe (young massive clusters, YMCs). SCs are found in a broad range of galactic environments and there are indications that the majority of clusters in the early and in the local Universe might have formed by the same mechanism. 

This scenario finds strong support in the consistency of the initial distribution of the clusters' masses (both in the early and in the local Universe) with an universal initial cluster mass function (ICMF), which is usually represented as a truncated power-law or Schechter distribution \citep{Zwart10,Kruijssen12}. Furthermore, the mass function of old cluster systems is found to be well represented by a log-normal distribution, the properties of which are almost independent of the host galaxy size, mass and morphological type \citep{Parmentier08,Harris14}. According to the results of the pioneering work of \cite{Vesperini97} and \cite{Vesperini98}, a log-normal distribution of the masses can be interpreted as the result of the evaporation of star clusters at the low-mass end of the mass function. However, as shown in \cite{Rossi15b} and in agreement with the results of \cite{Parmentier07}, there is a ``degeneracy'' in the ICMF, in the sense that both an initial log-normal and power-law mass spectrum can result in the observed bell-shaped distribution, at least for the case of the Galactic globular cluster system. On the other hand, \cite{Kruijssen12} showed that an initial log-normal ICMF in dwarf galaxies is inconsistent with the observed globular cluster systems, concluding that the initial distribution of the clusters' masses most likely follows a Schechter-type function of YMCs. In order to evaluate whether an ``universal'' power-law/Schecther initial mass function can evolve into a log-normal mass function almost independently of the properties of the host galaxy, it is thus important to understand the role played by the environment in shaping the evolution of star cluster systems (SCSs).

The impact of the galactic environment on the dynamical evolution of SCs has been studied in the past, both from an observational and a theoretical  perspective \citep[e.g.][]{Gnedin97,Baumgardt03,Gieles06,Gieles07,Hurley08,Lamers10,Kupper10,Kruijssen11,Gieles11,Smith13,
Renaud11,Berentzen12,Sanchez12,Rieder13,Silva-Villa14,Brockamp14,Renaud15b,Renaud15}. In several of these works the host galaxy has been modelled by using static (or rigid-body rotating) analytic parametric mass models. However, this simplistic approach finds a major limitation in the evidence that galaxies are dynamical systems evolving with cosmic time (for example the development of transient bars and spiral arms) and that can experience merging and tidal interaction within time scales typically shorter than the life-time of a star cluster (more than 10 Gyr for the old Galactic globular clusters, as a reference). A sophisticated modelling of the external tidal field is then required in order to self-consistently model the co-evolution of star clusters and galaxies.

Numerical simulations of galaxies are the best candidate to provide the require detailed information on the tidal field experienced by a star cluster, suggesting the need for a method allowing the coupling of $N$-body galaxy simulations with $N$-body star cluster simulations. However, this approach has to face the problems related  to the extremely different scales of evolution, both in time and in space, characterising  these dynamical systems. A solution was first proposed by \cite{Fujii07}, in which the authors describe an algorithm (\textsc{bridge}) designed to directly couple $N$-body simulations of star clusters in their parent $N$-body galaxy. For example, \cite{Rieder13} used a \textsc{bridge}-like scheme implemented in \textsc{amuse} \citep{Zwart11} to simulate the evolution of star clusters in a cosmological tidal field. The main advantage of this approach is that it allows the direct modelling of dynamical friction. On the other hand, as discussed by the authors, amongst the limitations of their method are the use of a softening length in the $N$-body simulations of the clusters, the assumption of equal-mass of the stars and the neglect of stellar evolution effects. An alternative solution based on a tensor approach (\textsc{nbody6tt}) has been proposed by \cite{Renaud11} and extended in \cite{Renaud15}. This method has been applied, for example, in \cite{Renaud13} to estimate the evolution of star clusters in merging galaxies. The main advantage of \textsc{nbody6tt} is that it is based on \textsc{nbody6} \citep{Aarseth03}, a state-of-the-art fully-collisional code to follow in great detail the evolution of collisional systems, such as star clusters, including a stellar mass function and stellar evolution effects.

Another solution to simulate star cluster evolution in a live galaxy is based on a parametrized cluster evolution method constructed through a tidal tensor approach \citep[e.g.][]{Kruijssen11}. The main advantage of this solution is that it doesn't require direct $N$-body simulation of star clusters. In fact, direct fully collisional $N$-body simulations of massive star clusters ($N\gtrsim 2\times 10^5$ stars) are computationally very expensive \citep{Heggie14}, posing limits on the capability to simulate the evolution of massive clusters. Thus the parametrised approach is particularly useful for estimating the impact of the galactic environment on a whole population of star clusters, where the clusters at the high mass end of the cluster mass function can not be directly modelled. Following this approach, in \cite{Rossi15b} and \cite{Rossi15a} (hereafter Paper I and Paper II) we have proposed a simple evolutionary model based on scaling relations calibrated to direct $N$-body simulations. This model proved to be valid to predict to a good approximation the dissolution time and mass-loss histories of star clusters evolving in a Milky Way-like galaxy modelled by using analytic mass models, both static and rigid-body rotating (central bar). 

In light of these results, the main goal of the present work is to quantify the effect of realistic galactic environments on the evolution of the properties of the hosted star cluster systems, with particular focus on the evolution of an universal initial mass function and its  dependence on the host galaxy characteristics. To achieve this goal we have developed an alternative method to couple $N$-body simulations of galaxies and of  clusters, in which the main idea is similar to the one at the base of the approach of \cite{Renaud11} and with the main difference that we overcome the approximation of linearised forces by reconstructing the tidal field around the star clusters. We then created a database of $N$-body simulations of galaxies with different masses and morphological properties and simulated the evolution of star clusters within these realistic potentials by using direct $N$-body simulations based on our new approach. By using the results of the simulations, we calibrated an evolutionary model to predict dissolution time, survival rate, minimum survival mass, radial density distribution and mass function evolution of star cluster systems in disc galaxies, dwarf galaxies and elliptical galaxies.

The paper is structured as follows. In Sec. \ref{sec:tidal_field_fit} we present the method adopted in the present work, including our approach to evolve clusters in realistic tidal fields and the database of $N$-body simulations of galaxies and clusters. Sec. \ref{sec:scaling_reliations} includes a discussion of an evolutionary model to predict dissolution times and masses of star clusters in realistic tidal fields. The evolution of the properties of star cluster systems in the different galaxy models, including the mass function, survival rates and density distribution is presented in Sec. \ref{sec:results}. We discuss our results in Sec. \ref{sec:discussion} and summarize our conclusions in Sec. \ref{sec:summary}.

\section{Methods}
One of the major goals of the present work is to develop a method that allows us to couple $N$-body simulations of galaxies with $N$-body simulations of star clusters. In our approach, we extract the information on the tidal field experienced by a star cluster from the galaxy-scale simulation, and use such information to define the galactic forces in the cluster-scale simulation. In this Section we first introduce the codes used to simulate the galaxies and the star clusters, and conclude with a description of the method developed for coupling the two codes.

\subsection{$N$-body galaxy simulations}
\label{sec:galaxy_simulations}
We investigate the dynamical evolution of disk galaxies with different masses and Hubble types in order to derive the time evolution of tidal fields of galaxies for the star cluster simulations. We adopt our original simulation code developed by \cite{Bekki13}, which enables us to investigate both (i) purely collisionless stellar systems (i.e., no gas and no star formation) (ii) systems composed of both gas and stars (with dissipative dynamics). Since the main purpose of this paper is to investigate the influences of global tidal fields of galaxies on GC evolution, we investigate the dynamical evolution of purely collisionless systems using the code.

Since the present  model for disk galaxies with different Hubble types is almost exactly the same as  that used in \cite{Bekki14}, we briefly describe the model in this paper. A disk galaxy  consists of a dark matter halo, stellar disk, and stellar bulge. The total masses of the dark matter halo, stellar disk,  and stellar bulge of a disk galaxy are denoted as $M_{\rm h}$, $M_{\rm d}$, and $M_{\rm b}$, respectively.

We adopt the density distribution of the NFW halo \citep{Navarro96} suggested from cold dark matter simulations in order to model the initial density profile of a dark matter halo in a disk galaxy. In this model we set $c=R_{\rm h}/R_{\rm s}=10$, where $R_{\rm h}$ and  $R_{\rm s}$ are the virial radius and the scale length of a dark matter halo, respectively. The bulge of a disk galaxy is represented by the Hernquist density profile. It has a size of $R_{\rm b}$ and a scale-length of $R_{\rm 0, b}$ ($=0.2R_{\rm b}$) and it is assumed to have an isotropic velocity dispersion. The bulge-to-disk ratio ($f_{\rm b}$) and $R_{\rm b}$ are set to be 0.167$M_{\rm d}$ and $0.2R_{\rm d}$, respectively, for a Milky Way (MW) type disk galaxy. In this notation $R_{\rm d}$ is the cutoff radius of the stellar disk. We also investigate models with different $f_{\rm b}$ (e.g., a pure disk galaxy has $f_{\rm b}=0$, see Tab. \ref{tab:database}).

The radial ($R$) and vertical ($Z$) density profiles of the stellar disk are assumed to be proportional to $\exp (-R/R_{0}) $ with scale length $R_{0} = 0.2R_{\rm s}$  and to ${\rm sech}^2 (Z/Z_{0})$ with scale length $Z_{0} = 0.04R_{\rm s}$, respectively. In addition to the rotational velocity caused by the gravitational field of the disk, bulge, and dark halo components, the initial radial and azimuthal velocity dispersions are assigned to the disc component according to the epicyclic theory with Toomre's parameter $Q$ = 1.5. The model with the parameters corresponding to the Milky Way (`MW') is regarded as the fiducial `MW' model in this study. The MW model is assumed to have $M_{\rm d}=6 \times 10^{10} {\rm M}_{\odot}$ ($=M_{\rm d, MW}$), $M_{\rm b}=10^{10} {\rm M}_{\odot}$ ($f_{\rm b}=0.167$), $R_{\rm d}=17.5$ kpc, $R_{\rm b}=3.5$ kpc, $c=10$, and $M_{\rm h} =10^{12} {\rm M}_{\odot}$. In our MW fiducial model we only included a galactic thin disc,  neglecting the contribution of an exponential thick disc \citep[e.g.][]{Juric08}. The maximum circular velocity of the MW model is set to 220 km/s which is consistent with the observed value \citep{Binney08}. 
The parameters of the dark matter halo model are consistent with lambda-CDM predictions \citep[e.g.][]{Neto07}. Furthermore, in lower-mass galaxies the mass ratio of dark matter to baryons can be higher \citep[see Fig. 18 of][]{Papastergis12} and we have indeed considered this in our set of models (see Secs. 3 and 4).

The total number of particles used in the fiducial model is $1\,083\,500$, and the mass and size resolutions of the stellar disk of the  model  are $1.2 \times 10^5 {\rm M}_{\odot}$ and 148 pc, respectively. The gravitational softening length ($\epsilon$) for a disk is chosen such that  $\epsilon$ can be the same as the mean particle separation at the half-mass radius of the disk. Therefore, $\epsilon$ can be different in different models with  $M_{\rm d}$ and $R_{\rm d}$. We mainly investigate 12 models with different $M_{\rm d}$, $R_{\rm d}$, and  $f_{\rm b}$, and the model parameters are given in Table 1.  The models without a stellar disk ($M_{\rm d}=0$) are referred to as elliptical galaxies,  whereas the model with the stellar disk having a similar mass as the LMC is referred to as the LMC model.

\subsection{$N$-body star cluster simulations}
\label{sec:star_cluster_simulations}
The simulations of the star clusters are performed with \textsc{nbody6} \citep{Aarseth03}. The code includes procedures to simulate the evolution of multiple systems, close encounters and collisions. Effects of single stellar evolution \citep{Hurley00} and binary stellar evolution \citep{Hurley02} are also included. The clusters in this study are initially in virial equilibrium, following a Plummer sphere distribution \citep{Plummer11}. The fraction of primordial binaries has been set equal to 5 per cent of the initial number of stars, while the binary orbital set-up has been chosen as described in \cite{Geller13}. A Kroupa mass function \citep{Kroupa01} is assumed for the initial distribution of the stellar masses. We neglected any dependence of the metallicity on the location of the clusters in the galaxy, assuming a value of $[\mathrm{Fe/H}]=-0.5$ for all the models. 

Velocity kicks imparted to neutron stars and black holes at birth in a supernova are uncertain (see Chatterjee, Rodriguez \& Rasio 2016 for a discussion) so we simply apply a kick chosen at random from a uniform distribution between 0 to 100 km/s. This results in about 5 neutron stars and black holes retained on average in the model clusters at 100 Myr, after the supernova phase has ceased. The escape velocity from clusters of larger-$N$ will be greater 
than for our $N = 10$k star clusters which means that the larger clusters will be expected to retain more remnants. Owing to mass-segregation 
these remnants will reside preferentially in the cluster core and can influence cluster properties \citep[e.g.][]{Breen13, Contenta15}. 
However, \cite{Chatterjee16} have used a series of large-$N$ Monte Carlo cluster models to show that the retention fraction
of black holes has only a minimal effect on the long term evolution of the total mass and dissolution time of a cluster.

We neglected any dependence of the initial size of the star clusters on their location in the galaxy, and chose Plummer models with initial half mass radius $r_\mathrm{hm} = 1.15$ pc. The impact of the choice of the initial size of star clusters is discussed in more detail in Sec. \ref{sec:initial_size_effect}. 

The simulated clusters are initially composed of 10k stars and the code allows us to follow their evolution until dissolution within typically one day of computing time on a single 16-core node with a NVIDIA Kepler K10 graphic processing unit (GPU).  In our approach these 10k stars simulations are used as a representation of larger clusters, as described in Sec. \ref{sec:scaling_reliations}. A technique to reduce the time required to complete the runs is to eliminate the stars that have escaped the cluster from the $N$-body simulation by defining an escape limit in \textsc{nbody6}. In order to include all the gravitationally bound stars it is fundamental to make sure that the escape limit in the simulations  is equal to or bigger than the maximum tidal radius. We thus evaluated the tidal radius of the clusters as discussed in Sec. \ref{sec:mesh_size} and assumed the escape limit to be equal to the tidal radius at the apogalactic passage of the clusters. 

\subsection{Reconstructing the tidal field}
\label{sec:tidal_field_fit}
In the present approach, the tidal forcing experienced by a star cluster is estimated directly from the galaxy $N$-body simulation. The procedure followed to do this can be schematically summarised in the following fours steps:
\begin{enumerate}
\item We select a particle in the $N$-body simulation of the galaxy, that we assume to trace the orbit of the centre of mass of the star cluster.
\item For each time step of the galaxy evolution, we create a regular grid of $4\times4\times4$ equally spaced points centred on the selected particle and evaluate the three components of the gravitational field ($a_x,a_y,a_z$) at each of the points of the grid from the galaxy $N$-body simulation. 
\item For each time step, we interpolate the values of each component of the gravitational field at each point of the grid with a 3D polynomial function of degree $n$ of the form:
\begin{align}
a_x(x,y,z) &= \sum\limits_{i+j+k=0}^n a_{ijk} x^i y^j z^k \nonumber\\
a_y(x,y,z) &= \sum\limits_{i+j+k=0}^n b_{ijk} x^i y^j z^k \;\;\; .\\
a_z(x,y,z) &= \sum\limits_{i+j+k=0}^n c_{ijk} x^i y^j z^k \nonumber
\end{align}
In this formulation, the first time derivative of the gravitational acceleration (jerk), required by \textsc{nbody6} to solve the equations of motion by using the Hermite integration scheme, can be easily derived as:
\begin{align}
\dot{a}_x(x,y,z) &= \sum\limits_{i+j+k=0}^n a_{ijk} \dfrac{\mathrm{d} (x^i y^j z^k)}{\mathrm{d} t}\nonumber\\
\dot{a}_y(x,y,z) &= \sum\limits_{i+j+k=0}^n b_{ijk} \dfrac{\mathrm{d} (x^i y^j z^k)}{\mathrm{d} t} \;\;\; .\\
\dot{a}_z(x,y,z) &= \sum\limits_{i+j+k=0}^n c_{ijk} \dfrac{\mathrm{d} (x^i y^j z^k)}{\mathrm{d} t} \nonumber
\end{align}
\item At this stage, for each time step of the galaxy simulation we have obtained a set of coefficients ($a_{ijk},b_{ijk},c_{ijk}$) defining both the gravitational field and jerk within a cubic volume containing the star cluster, which can be used as an input for the star cluster simulation, performed with \textsc{nbody6}.  We note that the information of the tidal field is updated at every time step of the galaxy simulation, for which the minimum value for our galaxy $N$-body code is 1.4 Myr. Considering that such a time step is considerably shorter than the typical time scales of dynamical evolution of our simulated star clusters (as a reference, the typical value of the relaxation time is of the order of a few hundreds Myr), the effects of a small discreteness of the time variation of the tidal field can be confidently neglected and, if necessary, it can always be altered.
\begin{figure}
\hskip -5mm
\includegraphics[scale=0.65]{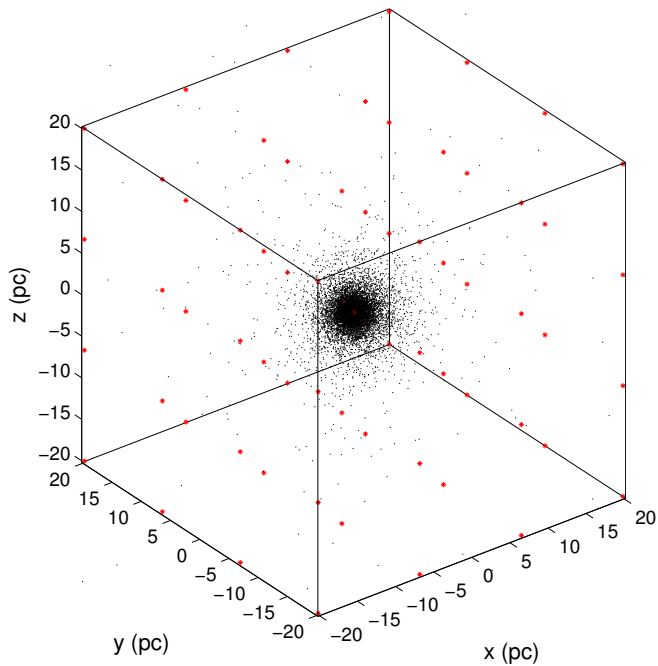}
\caption{Graphic visualisation of the method applied to sample and reconstruct the tidal field. The box shows the limits of the considered space volume centred on the cluster, the red points show the knots of the grid on which the gravitational field is evaluated from the galaxy $N$-body simulations. The star cluster is represented by the black points.}
\label{fig:grid}
\end{figure}

As a next step, we designed an experiment to test the level of accuracy of the  method. We selected three particles from an $N$-body model of the Milky Way galaxy following orbits at different galactocentric distances. The projection of the orbits on the galactic plane is shown in the top panel of Fig. \ref{fig:acc_test}. For each orbit, we evaluated the mean relative interpolation error for an $n=2$ degree polynomial model on the grid at each time step and compared the results for the three different orbits. The results of this analysis are shown in the bottom panel of Fig. \ref{fig:acc_test}. We note that, on average, the relative fitting error is smaller for orbits at larger distances from the galactic centre and increases in the inner regions of the galaxy. The reason for this behaviour is that the curvature of the gravitational potential is greater in the inner regions and decreases for increasing galactocentric distances. As a consequence, the polynomial fit reproduces the tidal field more accurately in the peripheral regions of the galaxy than in the inner regions. A possible solution to reduce this effect would be to increase the degree $n$ of the polynomial fitting function, which would on the other hand increase the complexity of the calculations, and hence the computing time. However, we note the relative error associated to a $n=2$ fit model is very small, being on average about $5.7\times 10^{-6}$, $6.3\times 10^{-6}$ and $6.7\times 10^{-6}$ for Orbit 1, Orbit 2 and Orbit 3, respectively. We can then conclude that a quadratic 3D polynomial fit can reproduce the gravitational field to a sufficient degree of accuracy.

\begin{figure}
\hskip -8mm
\begin{tabular}{r}
\includegraphics[scale=0.6]{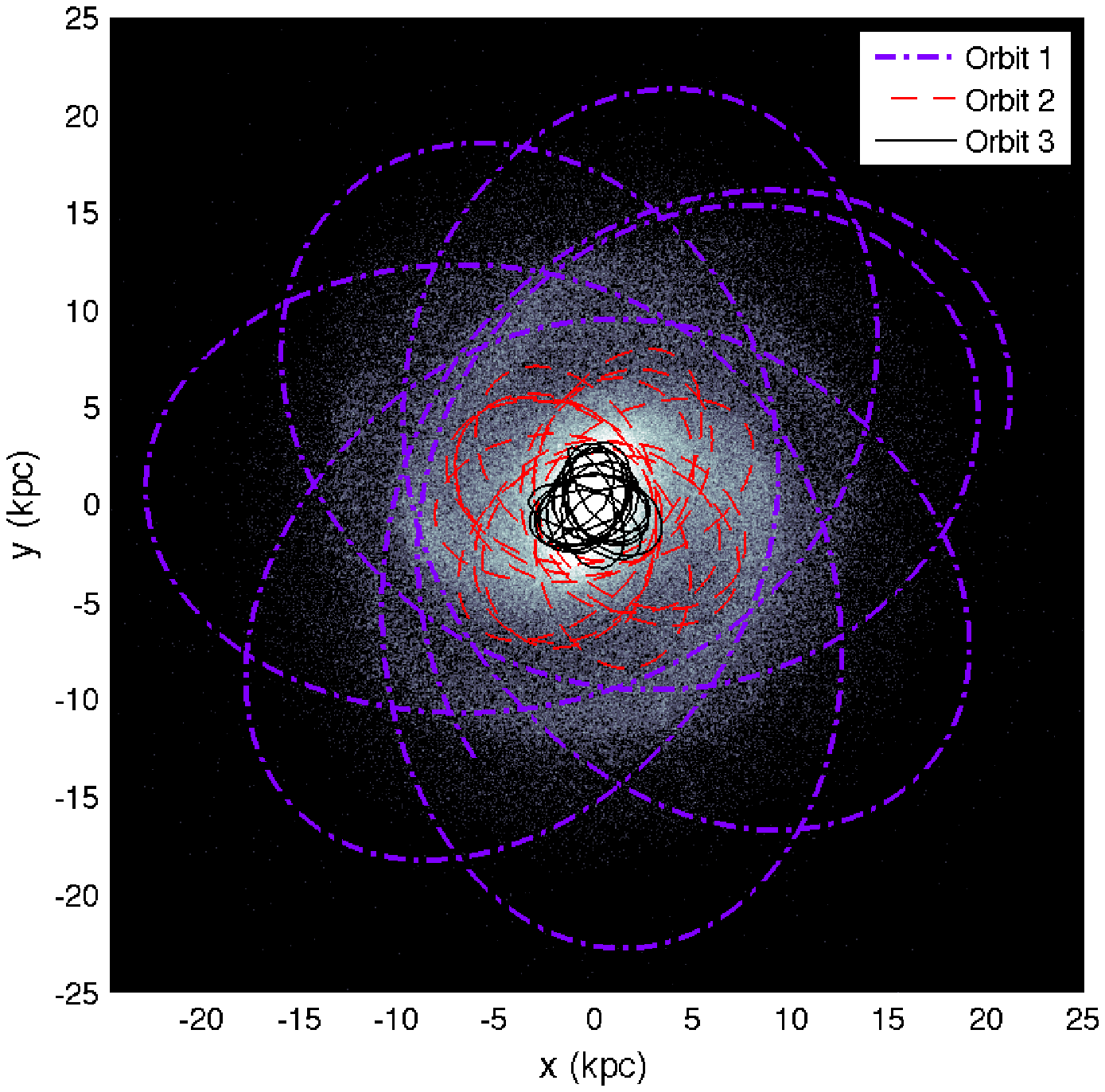}\\
\includegraphics[scale=0.61]{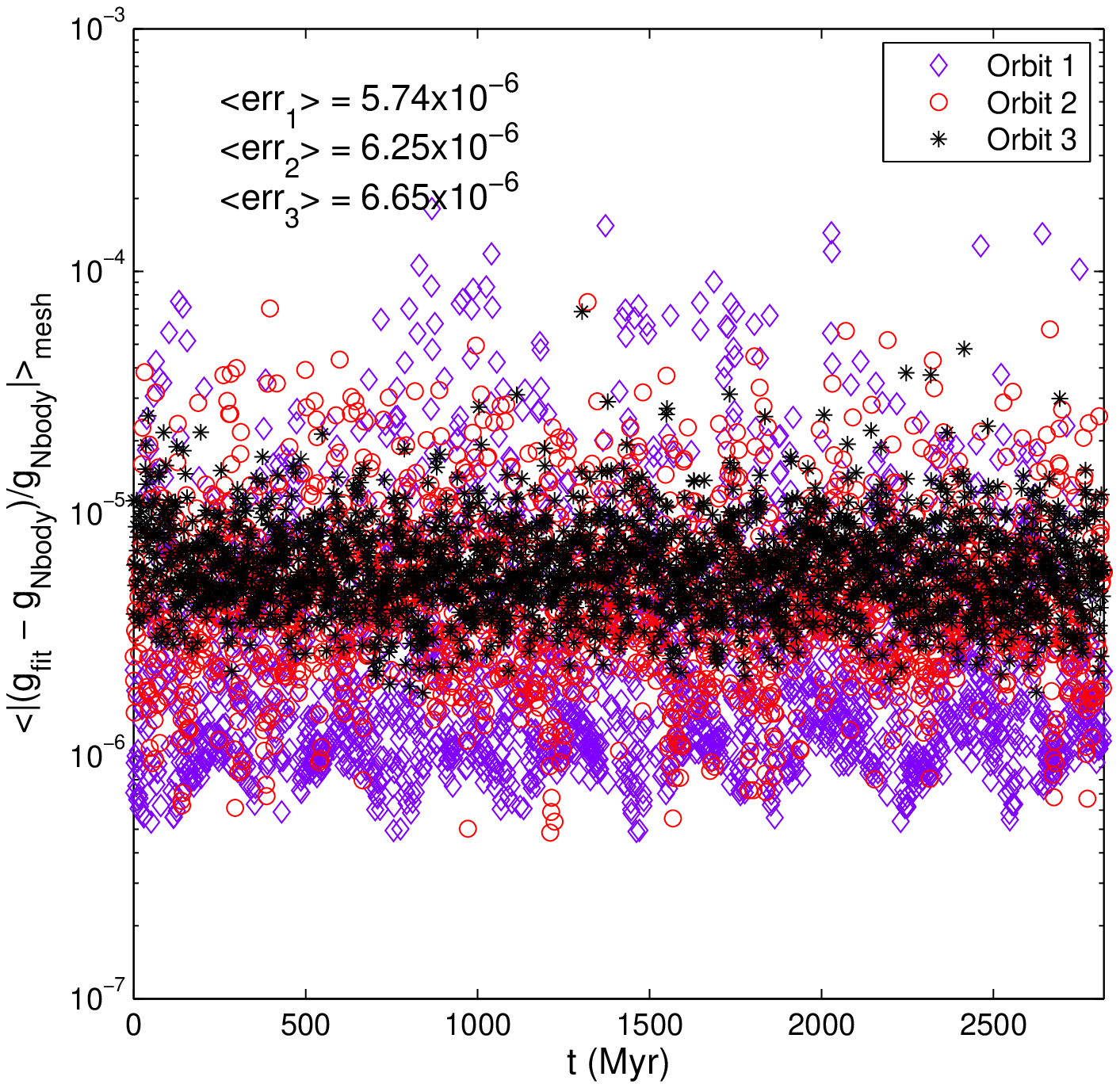}
\end{tabular}
\caption{\textit{Top panel}: representation of the orbits selected to test the accuracy of the proposed method, over-plotted to the considered galaxy model. \textit{Bottom panel}: time evolution of the relative error associated with the the polynomial fitting method with $n=2$ for the three different orbits considered. The vertical axis shows the relative difference between the value of the mesh-averaged gravitational field directly form the galaxy $N$-body simulation $a_\mathrm{Nbody}$ and the average value obtained from the 3D polynomial fitting procedure $a_\mathrm{fit}$ from eq. 1.}
\label{fig:acc_test}
\end{figure}

The main advantage of the present method is that, being based on \textsc{nbody6}, it is in principle applicable to the study of the evolution of star clusters with any structural/chemical parameters (initial mass, size, metallicity, etc.) on any orbit in any galactic tidal field (isolated disc, dwarf and elliptic galaxies, tidally interacting galaxies, galaxy clusters and cosmological tidal fields).
The main limitation of the current implementation is that it doesn't allow us to follow the orbits of the escapers once they have left the box sampling the tidal field. Such a restriction however doesn't affect our present analysis, which is focussed on the dynamical evolution of the gravitationally bound components of the clusters. A possible solution to overcome such a limitation would be to re-introduce the cluster escapers into the $N$-body simulation of the galaxy after they have left the box. This approach will be explored in future work. 
Another limitation of our approach is that it doesn't directly model the effects of dynamical friction. However, as we showed in Paper I, dynamical friction is expected to strongly affect the evolution of objects with the mass of a typical globular cluster clusters only in the very inner regions of the galaxy ($\sim 1.5$ kpc for a Milky Way-like galaxy). 
\end{enumerate}

\subsubsection{The size of the mesh}
\label{sec:mesh_size}
\begin{figure}
\hskip -7mm
\includegraphics[scale=0.53]{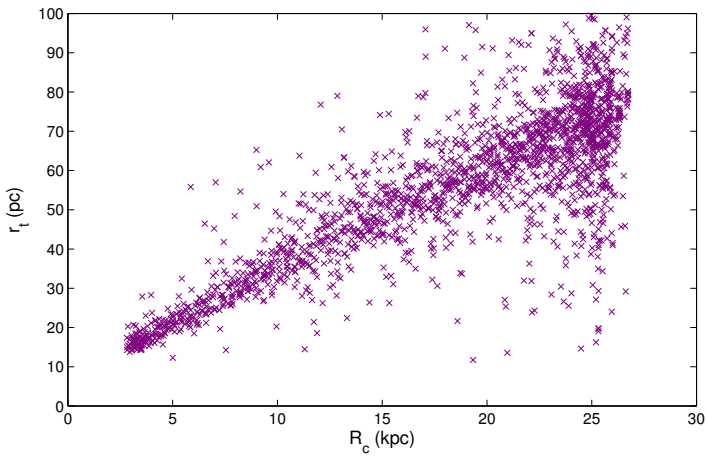}
\caption{Predicted tidal limit as a function of the galactocentric distance of a cluster composed of 10k stars and following an eccentric orbit in the GAL10 galaxy model.} 
\label{fig:rt_comparison}
\end{figure}

The size of the box within which the tidal field is sampled is a parameter that can be arbitrarily chosen in our approach. In the ideal scenario, the size of the box is big enough to contain the tidal radius of the star cluster to be modelled, and this would imply a dependence of the box size on the mass distribution of the host galaxy, on the mass of the star cluster and on its galactic orbit. We briefly recall that the escaped limit of a star cluster is defined as the distance from the cluster centre at which a star experiences an acceleration along the line connecting the cluster with the centre of the galaxy which is zero with respect to the cluster centre \citep{King62}. This translates into the condition
\begin{equation}
\ddot{R}_\mathrm{s} - \ddot{R}_\mathrm{c} = 0 \;\;\;,
\label{eq:tidal_radius}
\end{equation}
where 
\begin{align}
\ddot{R}_\mathrm{c} &= a_\mathrm{gal}(R_\mathrm{c}) \\
\ddot{R}_\mathrm{s} &= a_\mathrm{gal}(R_\mathrm{s}) - \dfrac{G M_\mathrm{c}(R_\mathrm{s} - R_\mathrm{c})}{|R_\mathrm{s} - R_\mathrm{c}|^3}\;\;\;.
\end{align}
In this notation $a_\mathrm{gal}$ is the projection along the position vector of the acceleration associated to the tidal field (including both the potential gradient and the centrifugal, Coriolis and tidal component), $R_\mathrm{c}$ and $R_\mathrm{s}$ are the galactocentric distance of the cluster centre and of a cluster star, respectively, and $M_\mathrm{c}$ is the mass of the cluster. By using the information on the tidal field from the polynomial interpolation method, we can predict the instant value of the tidal limit along the orbit of a cluster.  Fig. \ref{fig:rt_comparison} shows the instant value of the tidal radius obtained for a cluster with 10k stars orbiting within the GAL10 model (see Tab. \ref{tab:database}) by applying the definition in eq. \ref{eq:tidal_radius}. 
In the particular case shown in Fig. \ref{fig:rt_comparison}, the tidal radius of the cluster fluctuates between about 10 pc and 100 pc during its orbit (without considering any mass loss), and the ideal size of the mesh would be greater than the tidal radius at the apogalactic. However, we compared the estimated tidal field along the orbit of a star clusters by adopting meshes with different sizes, in particular 40 pc, 80 pc and 120 pc. In all the three test cases we found the same values of the reconstructed gravitational field, even at distances form the centre up to the maximum tidal limit bigger than the mesh size itself. This holds as an indication that even for mesh sizes smaller than the tidal limit of the star clusters the value of the gravitational field  can be extrapolated to a good accuracy outside the mesh up to distances comparable to the maximum tidal radius of the clusters. We then chose 40 pc as a ``standard'' size of the mesh for all the $N$-body simulations of the star clusters.

\subsection{The database}
\begin{figure*}
\includegraphics[scale=0.57]{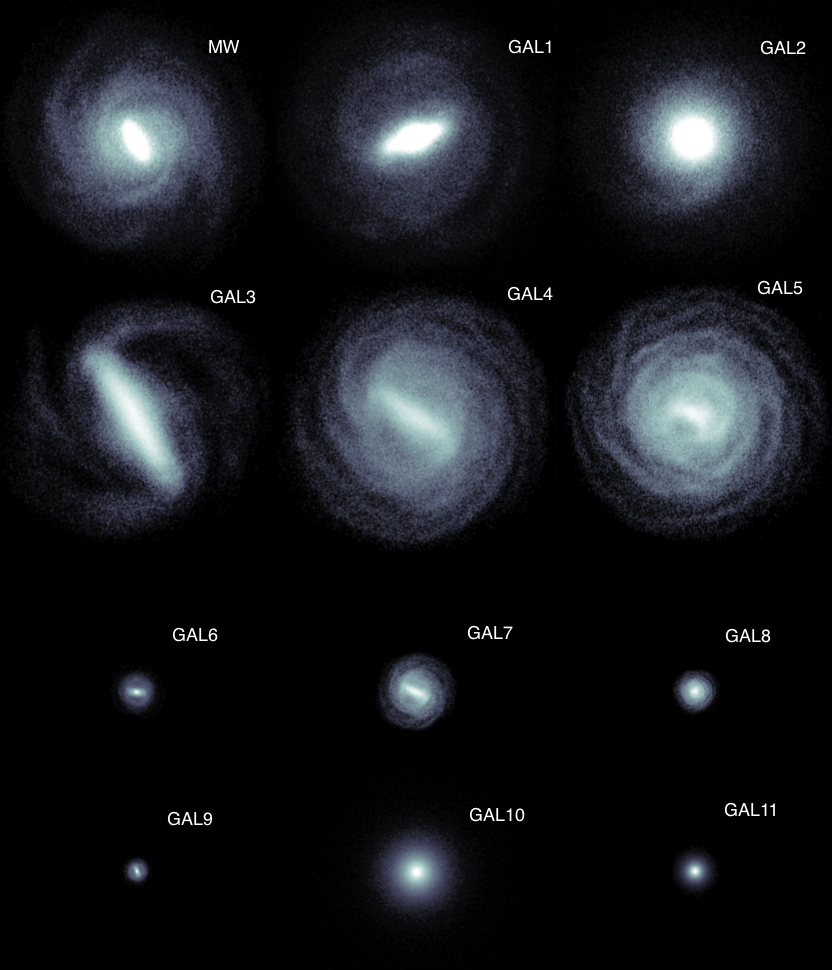}
\caption{Visualization of our database of galaxy models. From left to right and from top to bottom: MW, GAL1, GAL2, GAL3, GAL4, GAL5, GAL6, GAL7, GAL8, GAL9, GAL10, GAL11. The galaxies are represented using a logarithmic colomap. The relative size of the galaxies is the same. As a reference, the radius of the disc of the MW model is 17.5 kpc.}
\label{fig:galaxy_models}
\end{figure*}

Using the approach described in Sec. \ref{sec:galaxy_simulations} we generated a database of representative galaxy models. The sample includes six disc galaxies, four Magellanic-type galaxies and two elliptical galaxies. The main parameters of the various galaxy models are summarized in Tab. \ref{tab:database}, and are visualized in  Fig. \ref{fig:galaxy_models}. For each galaxy model, we selected eight particles, which are assumed  to trace the orbit of the star clusters, initially located at different distances from the centre of the galaxy. The initial locations of the eight particles in each model are randomly selected in order to uniformly cover distances and orbital eccentricities up to the maximum radial extent of the stellar component of the galaxy.
For each of the selected particles in each of the galaxy models, we reconstructed the tidal field following the method described in Sec. \ref{sec:tidal_field_fit} and used the information as an input for the simulations of the star clusters. This gives 96 distinct galaxy/orbit instances and for each of these we evolved a representative star cluster using the approach described in Sec. \ref{sec:tidal_field_fit}. Specifically, we followed $N$-body simulations of 96 clusters, each with initially 10k, and used the information on the dissolution times to calibrate equations predicting the evaporation rates, as described in detail in Sec. \ref{sec:scaling_reliations}. As noted in Sec. \ref{sec:galaxy_simulations}, the galaxy simulations are run to 2.8 Gyr of evolution. The reason for this choice was motivated by the expectation that the majority of the simulated clusters with 10k stars would dissolve within this time scale. In practice, we found a few exceptions, which were the clusters in the outer haloes of the galaxies. In these cases, however, the mass evolution is particularly regular, as a consequence of the smooth gravitational potential in the outer haloes of our simulated galaxies. We then have been able to use the information from the 2.8 Gyr of evolution to extrapolate the mass evolution of such clusters until dissolution by applying eq. \ref{eq:mass_loss} (see Sec. \ref{sec:scaling_reliations} for a more detailed discussion).

\begin{table*}
\centering
\begin{tabular}{c c c c c c c c}
\hline
\hline
\\
Name & $M_\mathrm{b}$ $(M_\odot)$ & $M_\mathrm{d}$ $(M_\odot)$ & $M_\mathrm{h}$ $(M_\odot)$ & $R_\mathrm{b}$ (kpc) & $R_\mathrm{d}$ (kpc)  & $R_\mathrm{h}$ (kpc) & Model   \\
\\
\hline
\hline
\\
MW       & $10^{10}$  & $6\times10^{10}$ & $10^{12}$ & 2.0 & 17.5 & 245 & Milky Way-like galaxy\\ 
\\
GAL1    & $3\times10^{10}$  & $6\times10^{10}$ & $10^{12}$ & 2.0 & 17.5 & 245 & Intermediate mass bulge spiral galaxy\\
\\
GAL2  & $6\times10^{10}$  & $6\times10^{10}$ & $10^{12}$ & 8.6 & 17.5 & 245 & Massive bulge spiral galaxy\\
\\
\hline
\\
GAL3    & -  & $6\times10^{10}$ & $6\times10^{12}$ & - & 17.5 & 245 & Spiral galaxy with strong bar \\
\\
GAL4    & -  & $3\times10^{10}$ & $6\times10^{12}$ & - & 17.5 & 245 & Spiral galaxy with intermediate-strength bar\\
\\
GAL5    & -  & $1.2\times10^{10}$ & $6\times10^{12}$ & - & 17.5 & 245 & Spiral galaxy with weak/no bar formation\\
\\
\hline
\\
GAL6    & - & $9\times10^{8}$ & $3\times10^{10}$ & - & 3.0 & 41.7 & LMC-like galaxy   \\
\\
GAL7    & - & $9\times10^{8}$ & $3\times10^{10}$ & - & 5.4 & 41.7 & Low-surface brightness LMC  \\
\\
GAL8    & - & $3.6\times10^{8}$ & $3\times10^{10}$ & - & 3.0 & 41.7 & LMC with lower disc/halo mass  \\
\\
GAL9   & - & $1.2\times10^{8}$ & $10^{10}$ & - & 1.75 & 24.5 & Dwarf disc  \\
\\
\hline
\\
GAL10   & $6\times10^{10}$  & - & $10^{12}$ & 8.6 & - & 245 & Elliptical   \\
\\
GAL11   & $6\times10^9$  & - & $10^{11}$ & 2.7 & - & 77.4 & Small elliptical  \\
\\
\hline
\end{tabular}
\caption{Database of $N$-body galaxy models. ($M_\mathrm{b}$, $M_\mathrm{d}$, $M_\mathrm{h}$) and  ($R_\mathrm{b}$, $R_\mathrm{d}$, $R_\mathrm{h}$) express the mass and the maximum extent of the mass (initially) in the bulge, disc and halo, respectively.}
\label{tab:database}
\end{table*}

\section{Scaling relations for star clusters in realistic tidal fields}
\label{sec:scaling_reliations}
As highlighted in Sec. \ref{sec:introduction}, the present-day methods do not allow us to directly simulate the dynamical evolution of an entire star cluster system. In fact, while simulations of smaller clusters can be run within a reasonable time frame, more massive clusters still require months or years of computation \citep{Heggie14} to be directly modelled. In a recent work \cite{Wang16} pushed the limit of the maximum number of particles in a direct collisional $N$-body simulation, following the evolution of star clusters with initially a million stars to 12 Gyr by using multiple GPUs, but only for  low density models and still taking at least six months per model. Several solutions to this problem, based on predictions of mass-loss rates calibrated to results of $N$-body simulations, have been proposed \citep[e.g.][]{Lamers10,Kruijssen11}. Following this approach, in Paper I and Paper II we proposed an evolutionary model which has proven to be valid to predict to a good approximation the mass evolution of clusters living within simplistic analytical static and rotating tidal fields.  In this section we briefly summarize the main ideas at the base of our approach and test its  validity in the more general case of a realistic tidal field generated by an $N$-body galaxy.

\subsection{Dissolution time of star clusters in galaxies}
We recall that the dissolution time of a star cluster orbiting within the gravitational potential of a galaxy can be expressed in the most general case as an equation of the form
\begin{equation}
t_\mathrm{diss} = f_\mathrm{int}*g_\mathrm{ext} \;\;\; ,
\end{equation} 
where $f_\mathrm{int}$ and $g_\mathrm{ext}$ are functions describing the contribution of internal (two-body relaxation) and external (tidal interaction) effects, respectively. For example, in the case of a star cluster on a circular orbit within a static, axisymmetric galaxy the previous equation becomes
\begin{equation}
t_\mathrm{diss} = k \left[ \dfrac{N}{\log(\gamma N)}\right]^x \left[ \dfrac{1}{r}\dfrac{d\Phi(r)}{dr}-\dfrac{d^2\Phi(r)}{d r^2}\right]^{-1/2}\;\;\;,
\label{eq:t_diss}
\end{equation}
where $N$ is the initial number of stars, $\gamma = 0.02$ is the Coulomb logarithm  \citep{Baumgardt03}, $r$ is the galactocentric distance of the cluster, $\Phi$ is the galaxy gravitational potential and ($k,x$) are parameters that can be calibrated to $N$-body simulations of star clusters \citep[e.g. Paper I,][]{Baumgardt03}. We recall that the assumption of the factor $\gamma = 0.02$ in the Coulomb logarithm is true only for a steep \citep{Salpeter55, Kroupa01} initial mass function \citep[see][]{Henon75,Farouki94}. Furthermore, because of the logarithmic dependence of the dissolution time to gamma, the adoption of a fixed value of gamma is not expected to significantly alter the outcome of our simulations.

\begin{figure*}
\hskip -8mm
\begin{tabular}{c c }
\includegraphics[scale=0.5]{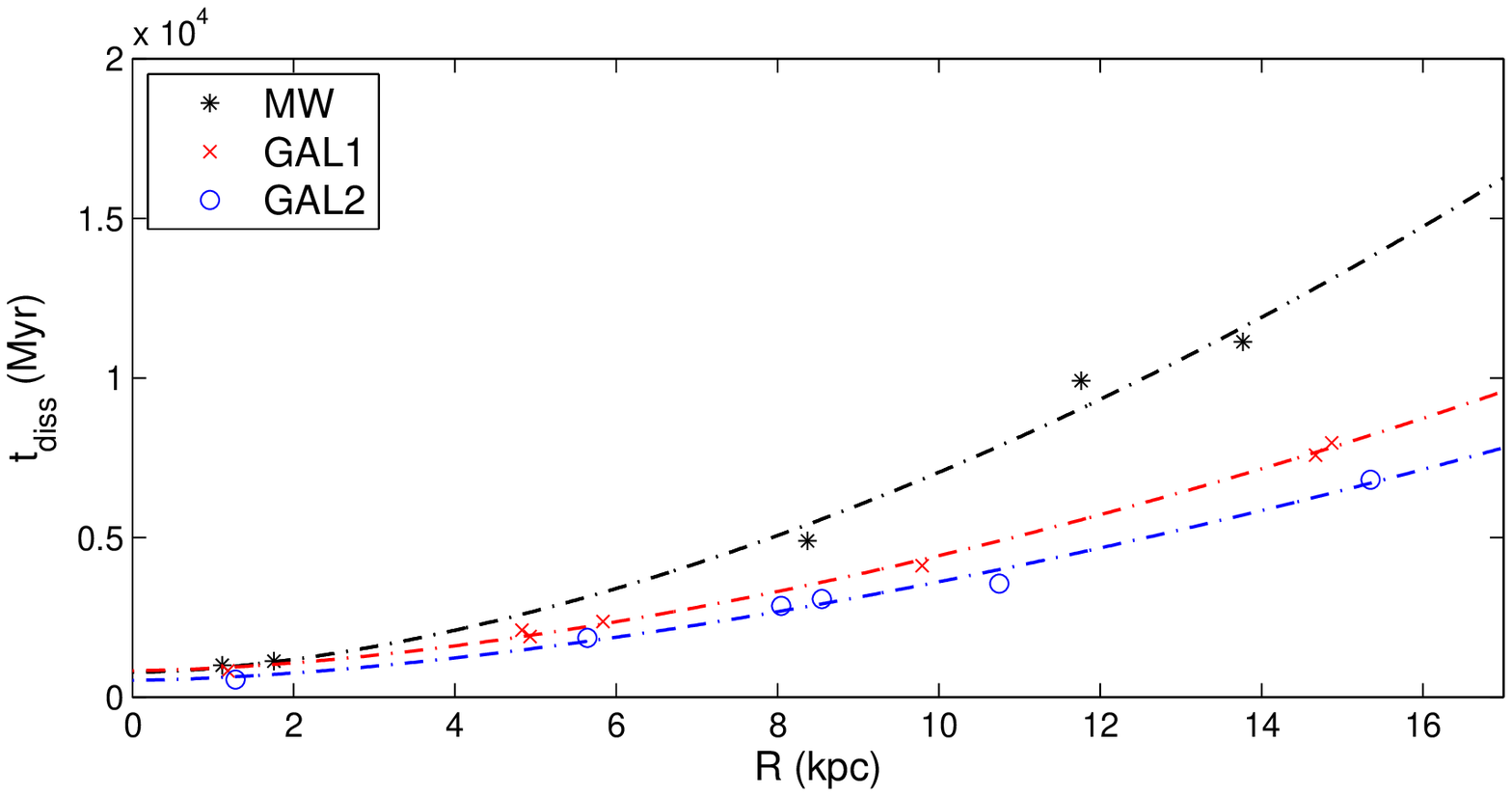} &
\includegraphics[scale=0.5]{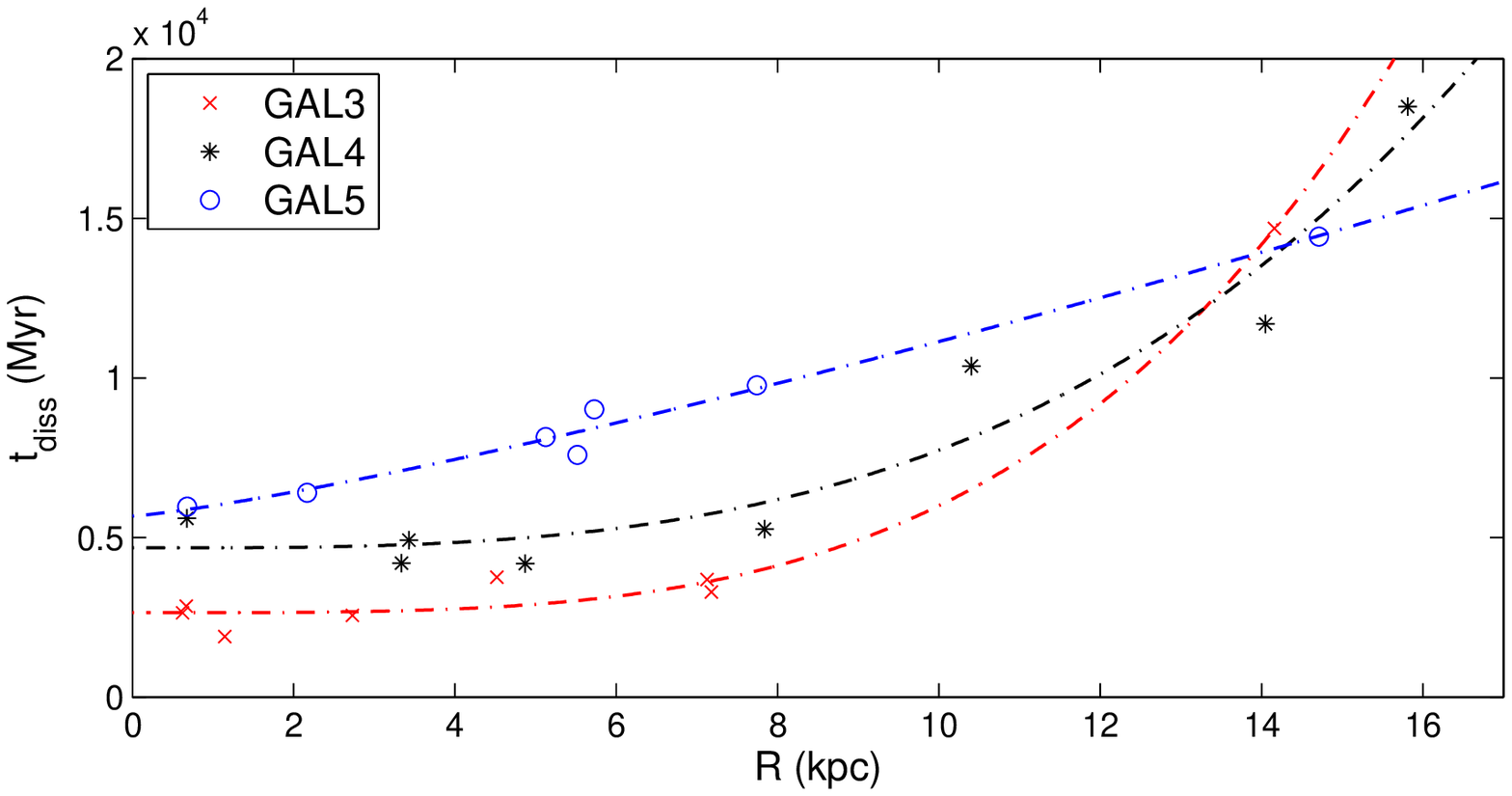} \\
\includegraphics[scale=0.5]{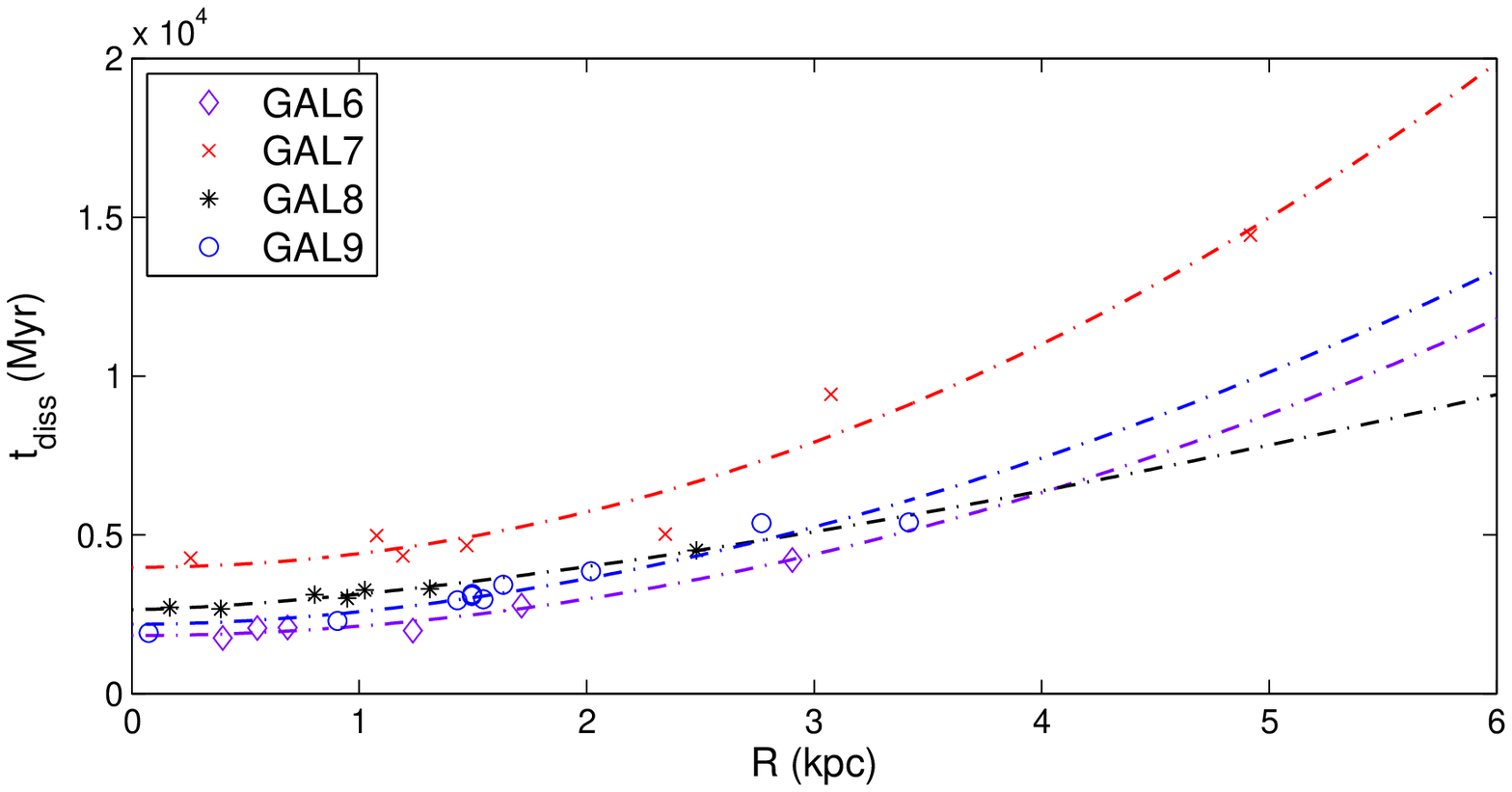} &
\includegraphics[scale=0.5]{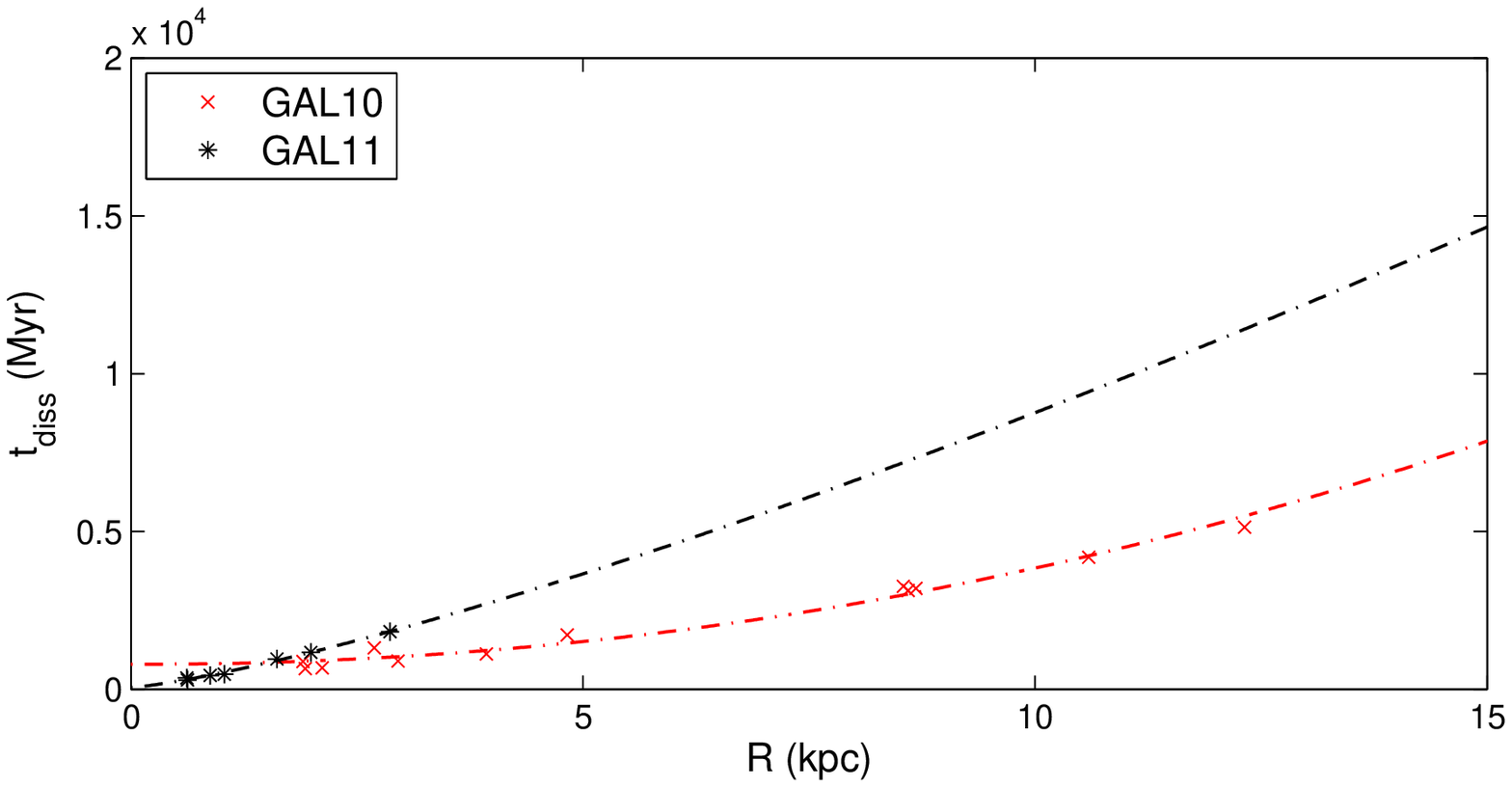} 
\end{tabular}
\caption{Dissolution time as function of the orbit-averaged galactocentric distance obtained from direct $N$-body simulations (points), interpolated with a power-law fitting function (lines). Each panel represents the results for galaxies grouped together based on their mass distribution and morphological type (see Tab. \ref{tab:database}).}
\label{fig:tdiss}
\end{figure*}

\begin{figure}
\hskip -5mm
\includegraphics[scale=0.5]{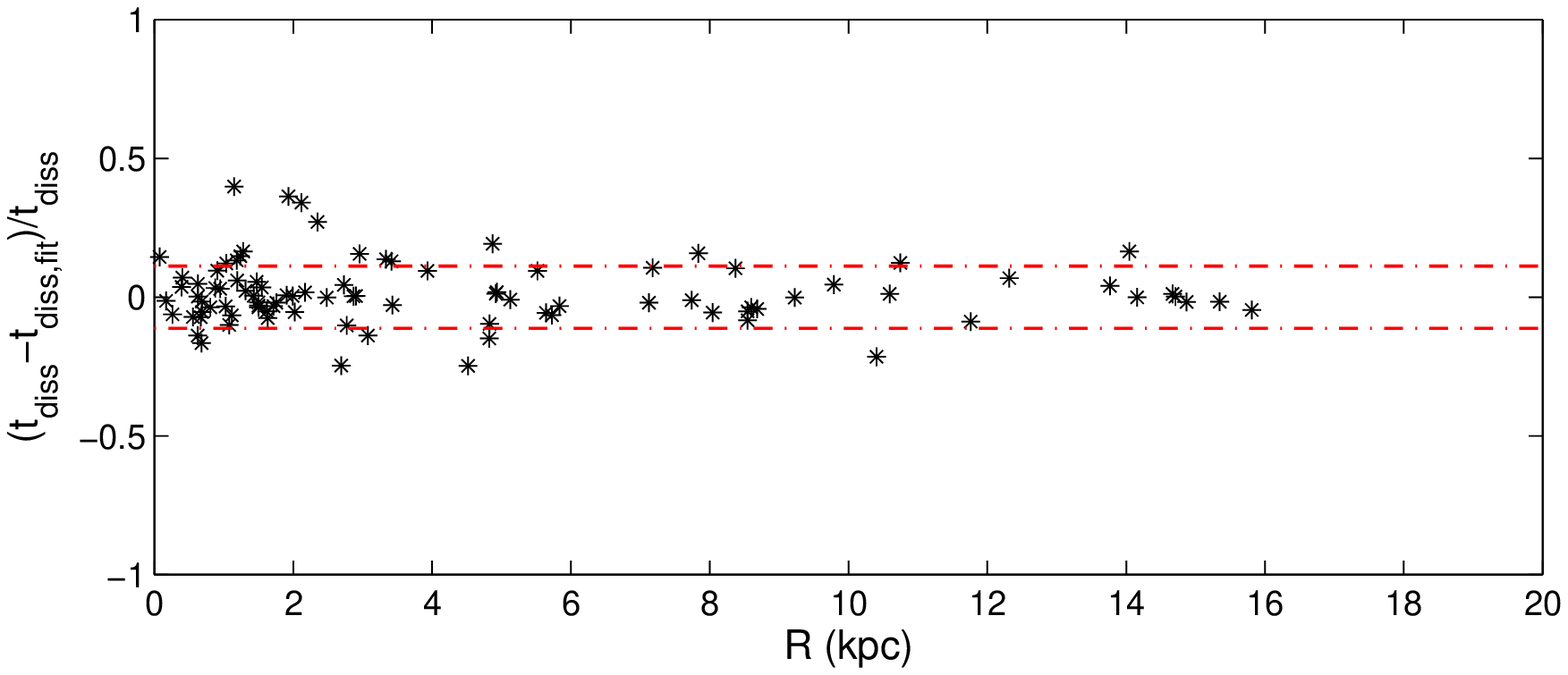}
\caption{Relative error associated to the estimate of the dissolution time as a function of the galactocentric distance for all the simulated clusters in the different models in our database (black points), obtained by using a power-law fitting function, whit parameters dependent on the galactic mass distribution. The red dot-dashed lines represent the standard relative error associated to the fit.}
\label{fig:tdiss_error}
\end{figure}

Eq. \ref{eq:t_diss} suggests that the dissolution time of a star cluster within a certain tidal field increases as a power-law function of the distance from the galactic centre $r$. However,  we note that $r$ is a constant value only in the specific case of circular orbits, while it varies during the orbital evolution for all the other cases. To take this effect into account, we considered a more general orbital angle-averaged galactocentric distance of the star clusters under a first approximation of elliptical orbits, defined as
\begin{equation}
R = R_\mathrm{p} \sqrt{\dfrac{1+e}{1-e}} \;\;\;,
\end{equation}
where $e = (R_\mathrm{a} - R_\mathrm{p})/(R_\mathrm{a} - R_\mathrm{p})$ and $R_\mathrm{a}$, $R_\mathrm{p}$ are the apogalacticon and the perigalacticon, respectively. Such an approach is in agreement with the results of \cite{Berentzen12}, in which the authors studied the dissolution of star cluster in a time-dependent tidal field generated by a rotating bar and found that the mass-loss of the cluster is mainly determined by the orbit-averaged tidal forcing. \footnote{We note that this result depends on the definition of ``average'' galactocentric distance. For instance, \cite{Webb14} used time-averaged galactocentric distances and found slightly different results.}
We then verified the validity of the power-law dependency in the case of more general tidal fields by using the results of our $N$-body simulations. Fig. \ref{fig:tdiss} shows the value of the dissolution times of clusters with initially 10k stars as a function of their orbit-averaged galactocentric distance for all the galaxy models in the database, which have been grouped based on their morphological characteristics and mass distributions (see Tab. \ref{tab:database}). 
For all the models, it appears that the dissolution times follow to a good approximation a simple power-law function of the orbit-averaged galactocentric distance $R$ of the form
\begin{equation}
t_\mathrm{diss} (R) = A + BR^{\alpha}
\label{eq:tdiss_general}
\end{equation}
where the parameters $A$, $B$ and $\alpha$ vary as a function of the galaxy properties. Their values obtained from a fit of the dissolution times (in Myr)  for cluster with initially 10k stars in the different galaxy models are summarized in Tab. \ref{tab:tdiss_parameter}. We also  note that the values of the coefficients in Tab. \ref{tab:tdiss_parameter} are obtained for initially compact clusters, and refer to Sec. \ref{sec:initial_size_effect} for more detailed discussion on the choice of the initial size of the clusters. Fig. \ref{fig:tdiss_error} shows the accuracy of the prediction of the dissolution times for all the galaxy models using a power-law fitting function. This power-law regression model allows us to derive the dissolution times of the stars clusters with standard relative percentage error equal to 11.2 \%, estimated by comparing the difference between the prediction of $t_\mathrm{diss}$ and results from $N$-body simulations for the whole sample of star clusters in our database. It is important to note that in our treatment the effect of disc-shocking is implicitly included in our approach, but its effect is averaged as we are considering orbit-averaged mass-loss rates. Strong disc shocking could be the effect responsible for the fluctuation of the observed dissolution times with respect to a power-law regression model, as suggested by the evidence that in the elliptical galaxy models the dissolution times of the star clusters follow a more regular trend (see bottom right panel of Fig. \ref{fig:tdiss}). However, we also note that disc shocking is not expected to be a dominant effect in modelling the evolution of the global properties of a whole population of star clusters \citep{Vesperini98}, unless low-$z$ orbits are considered. 

\begin{table} 
\centering
\begin{tabular}{c c c c}
\hline 
\hline
\\
Name & $A$ & $B$ & $\alpha$\\
\\
\hline
\hline
\\
MW    & $7.83 \times 10^2$  & $1.24 \times 10^2$   & 1.7\\
GAL1 & $8.32 \times 10^2$  & $0.76 \times 10^2$   & 1.7\\
GAL2 & $5.30 \times 10^2$  & $0.74 \times 10^2$   & 1.6\\
GAL3 & $2.65 \times 10^3$  & $0.07 \times 10^1$   & 3.7\\
GAL4 & $4.68 \times 10^3$  & $0.02 \times 10^2$   & 3.2\\
GAL5 & $5.68 \times 10^3$  & $3.23 \times 10^2$   & 1.2\\
GAL6 & $1.83 \times 10^3$  & $2.95 \times 10^2$   & 2.0\\
GAL7 & $3.98 \times 10^3$  & $4.32 \times 10^2$   & 2.0\\
GAL8 & $2.64 \times 10^3$  & $4.92 \times 10^2$   & 1.5\\
GAL9 & $2.19 \times 10^3$  & $3.95 \times 10^2$   & 1.9\\
GAL10 & $7.90 \times 10^2$  & $0.25 \times 10^2$   & 2.1\\
GAL11 & $0.57 \times 10^2$  & $4.64 \times 10^2$   & 1.3\\
\\
\hline
\end{tabular}
\caption{Parameters of the dissolution time power-law (eq. \ref{eq:t_diss}), defining the dissolution time (in Myr) of clusters with initially 10k stars as a  function of their orbit-averaged galactocentric distance and of the host galaxy model.}
\label{tab:tdiss_parameter}
\end{table}

Eq. \ref{eq:t_diss} also suggests that the ratio  of the dissolution times of two clusters of different initial masses $M_1$ and $M_2$ (or, equivalently, different initial number of stars $N_1$ and $N_2$) following the same orbit (i.e. experiencing the same gravitational tidal field) is a function of the initial mass only, the contribution of the tidal field disappearing as a common factor. More specifically, we can write
\begin{equation}
\dfrac{t_\mathrm{diss}(N_1)}{t_\mathrm{diss}(N_2)} = \left[ \dfrac{ N_1\ln (\gamma N_2)}{N_2 \ln(\gamma N_1)} \right]^x\;\;\; ,
\label{eq:tdiss_ratio}
\end{equation}
where $x=0.88$ (Paper I). The greatest advantage of this relation is that the dissolution time of a small cluster can be used to predict the dissolution time of a cluster with any initial mass following the same orbit, as discussed in Paper II. A potential issue in the use of eq. \ref{eq:tdiss_ratio} as a scaling relation is that some processes during escape depend on the number of paricles and on the escape criterion in a complex way \citep[see][]{Baumgardt01}. However, this effect is expect to be non-negligible for low-$N$ star clusters and only marginal for clusters with masses in the range considered in the present work.  In the next section we check the validity of this key result in the more general case of realistic tidal fields and show how it can be applied to predict the mass evolution of star clusters.

\subsection{Mass scaling relation}

In our previous work we have found that the mass evolution of a star cluster  can be well represented by an equation of the form
\begin{equation}
M(t) = M(0)\left[ 1-\left(\dfrac{t}{t_\mathrm{diss}}\right)^{\beta}\right] \;\;\;,
\label{eq:mass_loss}
\end{equation}
which represents a slightly modified version of the equation proposed by \cite{Baumgardt03}. In this case, the mass at any given time is determined by the dissolution time $t_\mathrm{diss}$ and the parameter $\beta$. As noted in Paper II, this equation is not able to reproduce the details of the small-scale mass-loss-rate fluctuations owing, for example, to bulge/disc shocks. On the other hand, we found that the overall trend of the mass evolution over the orbital time scale of the cluster in the galaxy follows eq. \ref{eq:mass_loss} to a good approximation.

\begin{figure*}
\hskip 15mm
\begin{tabular}{c c}
\begin{adjustbox}{valign=t}
\includegraphics[scale=0.6]{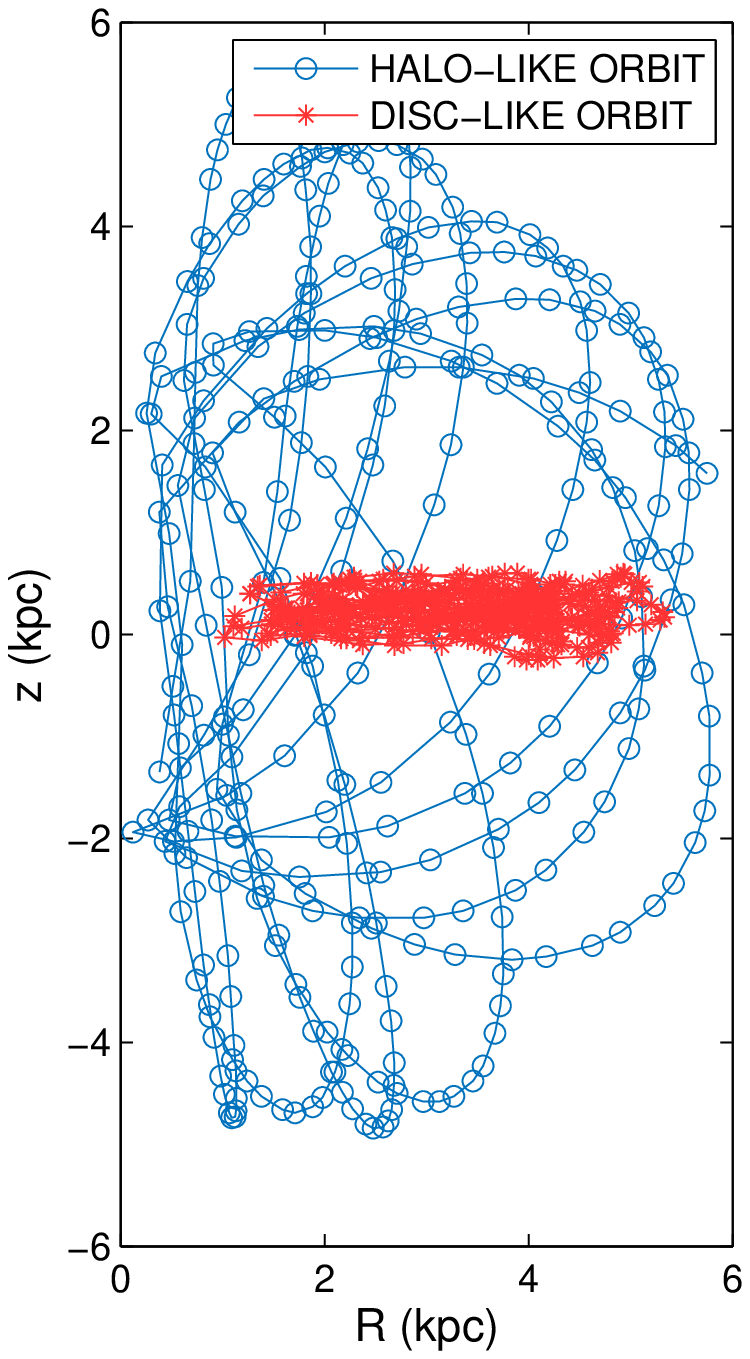}
\end{adjustbox}
&
\begin{adjustbox}{valign=t}
\begin{tabular}{@{}c@{}}
\includegraphics[scale=0.52]{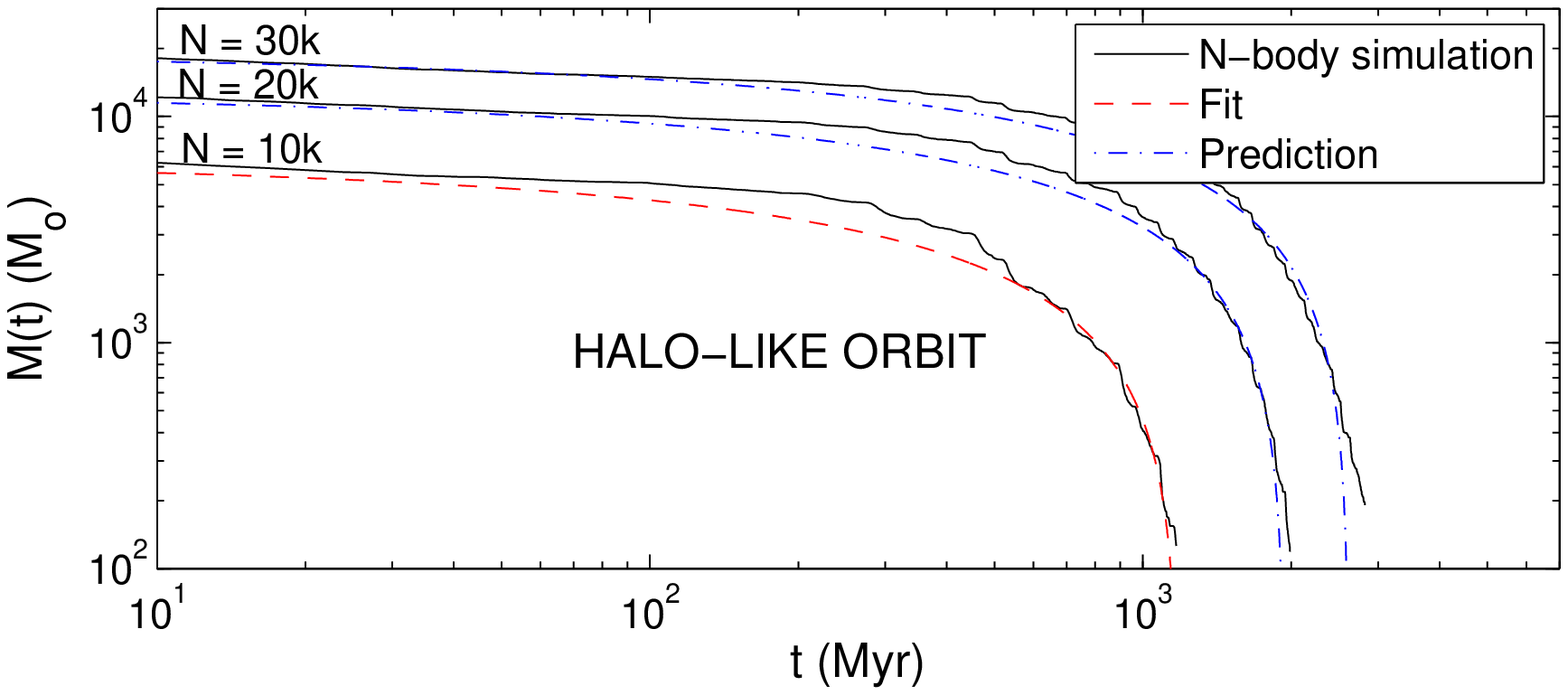} \\
\includegraphics[scale=0.52]{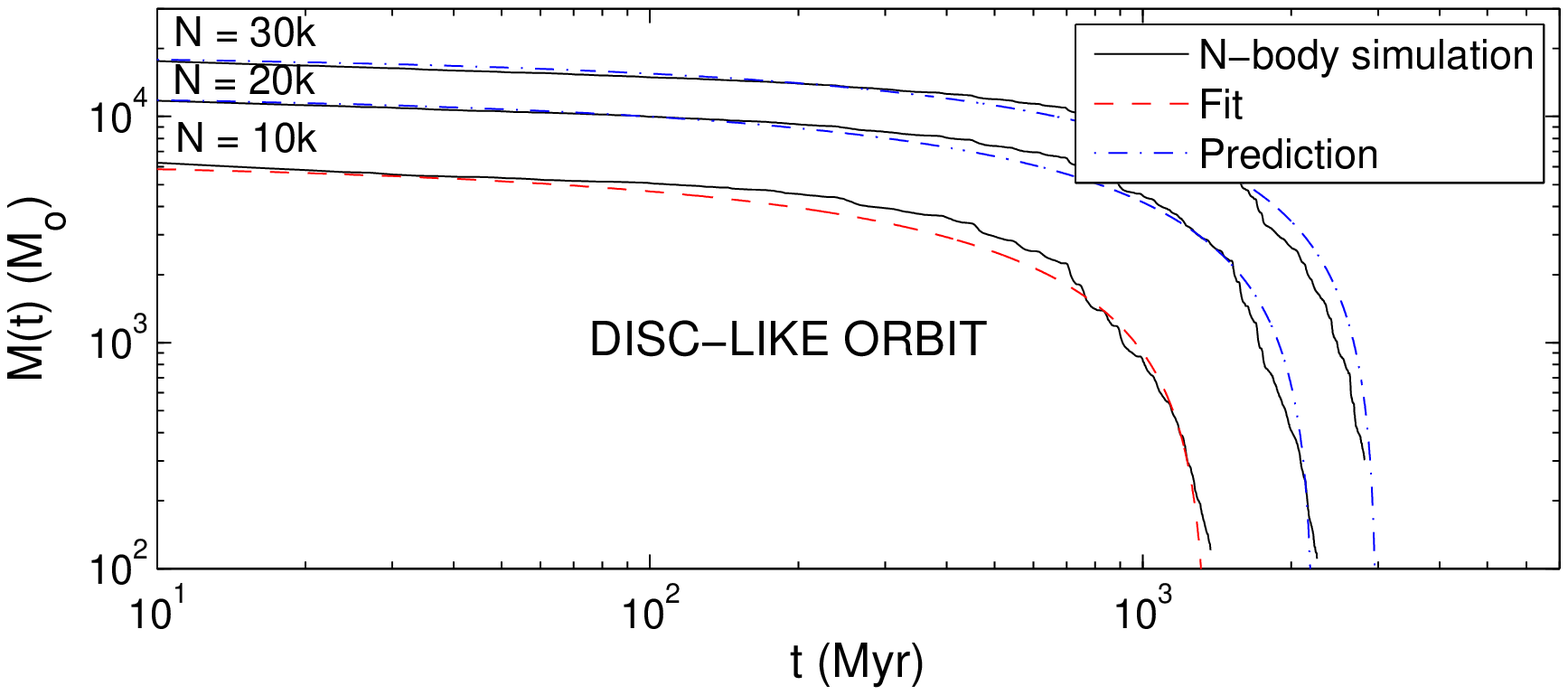}
\end{tabular}
\end{adjustbox}
\end{tabular}
\caption{\textit{Left panel}: Projection of the selected orbits in the meridional plane $(R,z)$. The blue open points trace the trajectory of the halo-like orbit, while the red stars trace the orbit of the disc-like orbit. \textit{Right panels}: Mass evolution of star clusters with different initial masses and following the selected halo-like orbit (top panel) and the disc-like orbit (bottom panel). The black lines show the results of direct $N$-body simulations. The red dotted lines represent the fit of the mass evolution with eq. \ref{eq:mass_loss} to the 10k simulation data and blue dot-dashed lines show the predictions of the evolutionary model for N = 20k and 30k based on the discussed scaling relations.}
\label{fig:test_scaling_relations}
\end{figure*}

\begin{figure}
\hskip -5mm
\includegraphics[scale=0.5]{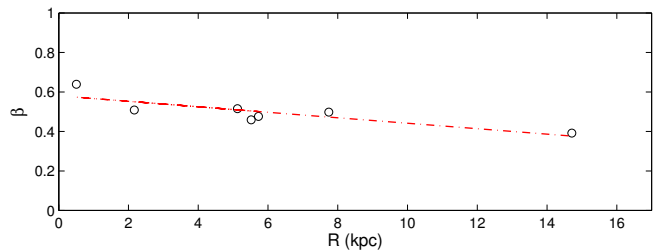}
\caption{Example of the dependence of the $\beta$ parameter of equation \ref{eq:mass_loss} on the galactocentric distance $R$ for the GAL5 model. The open black points show the results from $N$-body simulations, while the red dot-dashed line represents a linear regression interpolation model.}
\label{fig:beta_dependency}
\end{figure}

In order to test the validity of these results for the new description of the tidal field, we selected two particles in the MW galaxy model (see Tab. \ref{tab:database}) orbiting  within 6 kpc from the galactic centre. One of them describes a disc--like quasi-planar orbit with a mild inclination to the galactic plane, whereas the other one traces a trajectory included within a roughly spherical volume, which we can consider to be representative of an inner halo orbit (see Fig. \ref{fig:test_scaling_relations}). For each orbit, we ran a set of simulations of star clusters with initially 10k, 20k and 30k stars, respectively. We determined the value of $t_\mathrm{diss}$ and $\beta$ for the clusters with initially 10k stars for each orbit and applied the scaling relations  (eq. \ref{eq:tdiss_ratio} and \ref{eq:mass_loss}) to predict the mass evolution of the clusters with 20k and 30k stars. Fig. \ref{fig:test_scaling_relations} shows the results of this test, where the predictions of the model based on the 10k stars simulation are compared with the results of the direct 20k and 30k stars $N$-body simulations.  We can see that the overall mass evolution of the clusters is well described by the power-law function of eq. \ref{eq:mass_loss}. The dissolution times (not shown) also scale well, following eq. \ref{eq:tdiss_ratio}. Furthermore, the results of the $N$-body simulations are in good agreement with our predictions for both the disc and the halo clusters. We found that the  dependence of the mass evolution on the orbit is described by the parameter $\beta$. According to our results the value of $\beta$ scales with a very small linear dependence on the galactocentric distance $R$ of the cluster of the form
\begin{equation}
\beta = C + DR \;\;\; .
\label{eq:beta_dependency}
\end{equation}
We show an example in Fig. \ref{fig:beta_dependency}, where the value of $\beta$ as function of the galactocentric distance $R$ obtained for the galaxy model GAL5 has been fitted with a linear regression model. In other words, clusters closer to the galactic centre of their host galaxy experience a more sustained mass-loss owing to the more intense gravitational field, while clusters in more peripheral region are characterized by a reduced gradient of the mass-loss evolution. The value of the parameters $C$ and $D$ in eq. (\ref{eq:beta_dependency}) for each galaxy in the database are summarised in Tab. \ref{tab:beta_parameters}.
\begin{table} 
\centering
\begin{tabular}{c c c }
\hline 
\hline
\\
Name & $C$ & $D$ \\
\\
\hline
\hline
\\
MW    & 0.77 & -0.031 \\
GAL1 & 0.66  & -0.012\\
GAL2 & 0.66   & -0.008\\
GAL3 & 0.80  & -0.030\\
GAL4 & 0.71  & -0.020\\
GAL5 & 0.59  &   -0.013\\
GAL6 & 0.57   & 0.008\\
GAL7 & 0.77   & -0.082\\
GAL8 & 0.54   &  0.019\\
GAL9 & 0.67   & -0.041\\
GAL10 & 0.67  & -0.012\\
GAL11 & 0.73   &-0.071\\
\\
\hline
\end{tabular}
\caption{Value of the parameters of eq. (\ref{eq:beta_dependency}), defining the value of the gradient of the mass evolution of a cluster as function of its galactocentric distance. In these units, the galactocentric distance of the clusters $R$ is expressed in kpc.}
\label{tab:beta_parameters}
\end{table}

 The only two exceptions to this negative trend are found for the GAL6 and GAL8 Magellanic-type dwarf models, in which $\beta$ slightly shows a small positive dependency on the galactocentric distance. A more detailed analysis reveals a large scatter of the $\beta$ values for these galaxies, inducing a larger uncertainty of the value of the slope of eq. (\ref{eq:beta_dependency}). This could reflect the fact that the clusters in these dwarf galaxy models evolve quite close to the galactic centre, a region in which the $\beta$ dependency might not be well defined. Furthermore, when taking the error of the slope in eq. (\ref{eq:beta_dependency}) into account we found values consistent with a negative trend also for these two peculiar results.

Summarising the results of this section, we have found that the dissolution time of globular clusters in different galactic environments increases as a power-law function of their orbit-averaged galactocentric distance. Scaling relations describing dissolution time and mass evolution of the clusters, already proven to be valid in the more simplistic case of analytic gravitational potentials, are valid also in the case of $N$-body representations of isolated galaxies.  We have obtained an evolutionary model calibrated to small $N$-body star clusters simulations (10k stars) that can allow us to quantify the evolution of star cluster systems as a function of the galactic environment. This analysis is presented in the next section.

\section{The evolution of star cluster systems along the Hubble sequence}
\label{sec:results}
By using the fundamental scaling relations presented in Sec. \ref{sec:scaling_reliations} we have calibrated evolutionary equations predicting the value of the mass of a star cluster with any arbitrary initial mass and at any orbit-averaged position in the host galaxy, provided we have a reference simulation for the orbit of interest in the host galaxy of interest. The next step of our analysis is to quantify how different realistic galactic environments affect the evolution of the properties of their hosted star cluster systems. Particularly interesting for the subject of the present study, the method developed allows us to predict the evolution of the star cluster mass function, the expected survival rate of clusters as a function of the location in the galaxy and the minimum survival mass required for the clusters in order to survive a certain cosmic time.
However, the properties of an evolved star cluster system are strongly dependent on the initial conditions of the cluster system itself. For example the initial distribution of the masses of the clusters, the dependence of the initial masses on the location of the formation site of the clusters, initial number distribution in the galaxy and the age of the clusters are expected to affect the subsequent evolution of system. 
To proceed we will select a set of ``standard'' initial conditions of the star cluster systems for all the galaxies, similar to the estimated initial conditions of the Galactic globular cluster system. This way we can compare the evolution of systems with identical initial properties and gauge the dependence on the galactic environment.

\subsubsection*{Initial conditions of the star cluster systems}
We chose a ``universal'' initial cluster mass function described as a truncated power-law of the form
\begin{equation}
dN \propto M^{\gamma} dM
\end{equation}
where $\gamma = -2$. Following \cite{Brockamp14}, the overall mass range of the initial cluster population is chosen to be $M_\mathrm{cl} = 10^4-10^7$ $M_\odot$. As a first approximation, we neglected any dependency of the initial size of the clusters on their galactic coordinates. The effects of different choices of the initial sizes are discussed in more detail in Sec \ref{sec:initial_size_effect}. We assumed all the clusters to be coeval and, since we are studying isolated galaxies, we neglected any subsequent formation episode triggered, for example, by galaxy mergers and interactions. According to \cite{Vesperini98} and \cite{Parmentier05}, we distributed the initial number of clusters in each galaxy following a density profile characterising the Galactic halo globular clusters and scaling as
\begin{equation}
\sigma(R) \propto R^{-3.5}\;\;\;,
\end{equation}
where $R$ is the galactocentric distance of the cluster. We neglected any dependency of the initial mass of the clusters on their galactic coordinates. Consistent with the properties of the Galactic GC system, the radial extension of the star cluster system for the disc galaxy models is assumed to be twice the size of the stellar matter component of the host galaxy. In agreement with \cite{Kartha14}, the radial extent of star cluster systems in the elliptical galaxy models is set to be 14 times the effective stellar radius of the host galaxy. In our approach the effective radius has been defined as the scale length of the adopted Hernquist density profile (see Sec. \ref{sec:galaxy_simulations}). Finally, we assumed the size of the globular cluster systems in the dwarf Magellanic-type galaxies to be equal to the radius of the stellar component. Each cluster system is evolved for a time of 13 Gyr, which is roughly equal to the average age of the Milky Way GCs.  In order to obtain statistically significant results, for each galaxy we created a synthetic star cluster population composed of $10^5$ objects. \\

As a next step, we used our evolutionary model consisting of the prediction of the dissolution time as function of the galactocentric distance (eq. \ref{eq:t_diss}) for clusters with initially 10k stars, the scaling relation (eq. \ref{eq:tdiss_ratio}) and the mass evolution function (eq. \ref{eq:mass_loss}) to predict the mass after 13 Gyr of evolution of each one of the $10^5$ clusters of the generated populations for each of the galaxy models. This procedure allowed us to generate the evolved mass functions.

\subsection{Star cluster systems in disc galaxies with a massive bulge}
\label{sec:bulge_galaxies}
\begin{figure}
\centering
\includegraphics[scale=0.5]{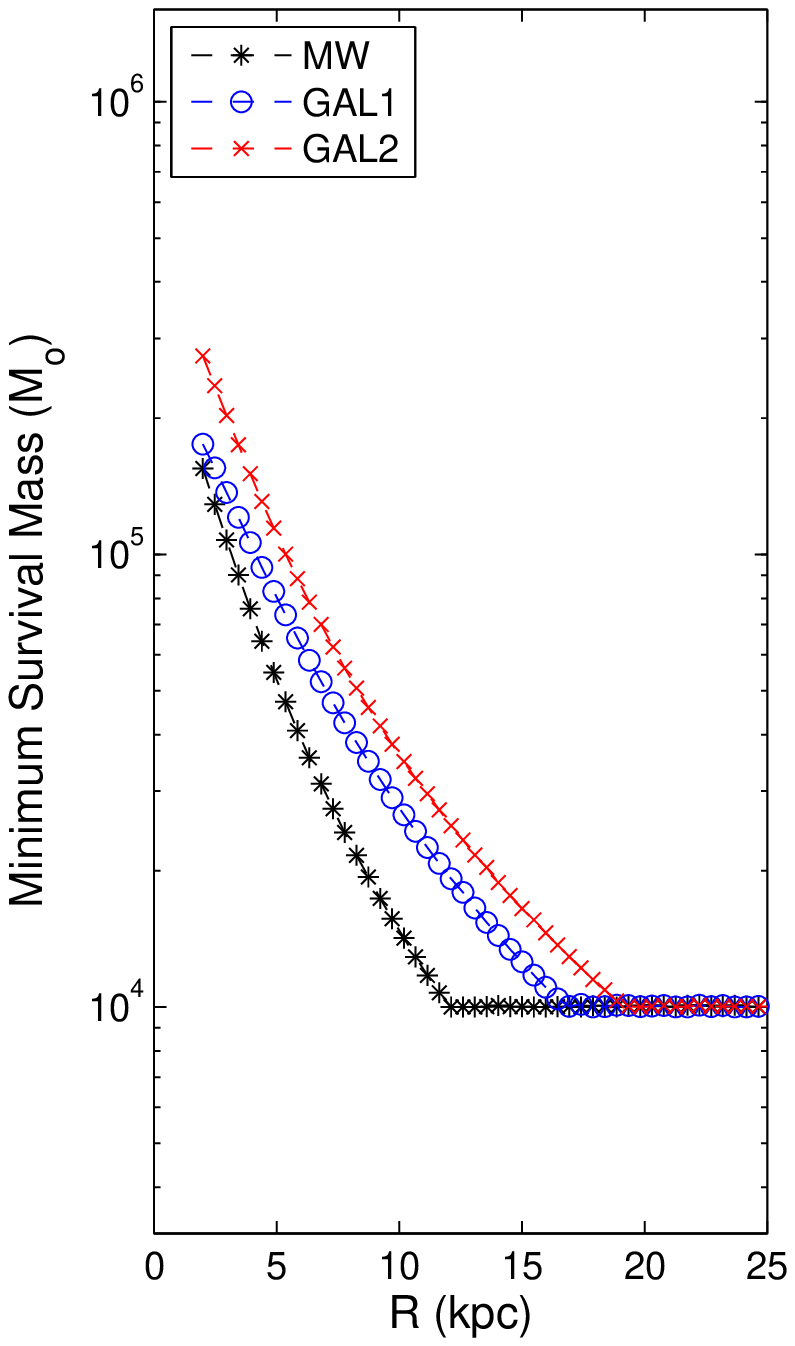}
\includegraphics[scale=0.5]{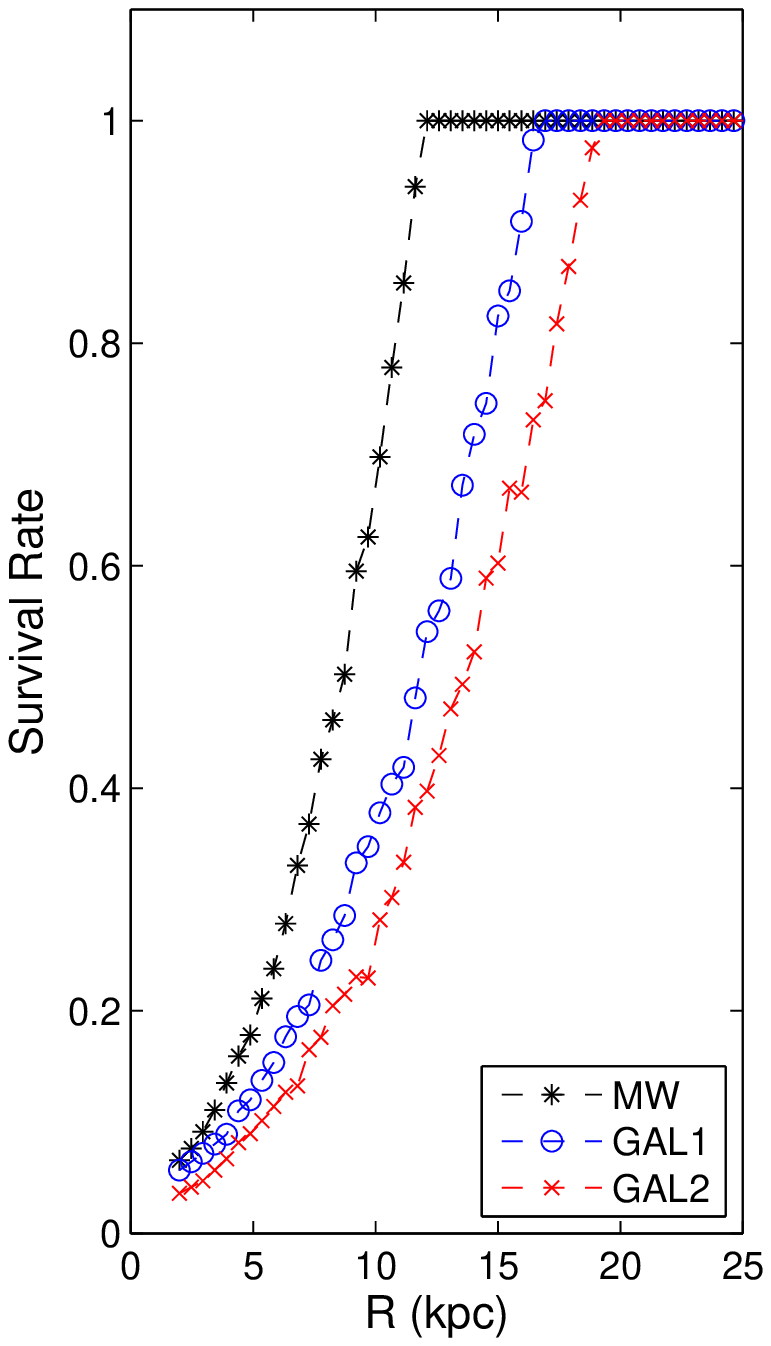}\\
\includegraphics[scale=0.58]{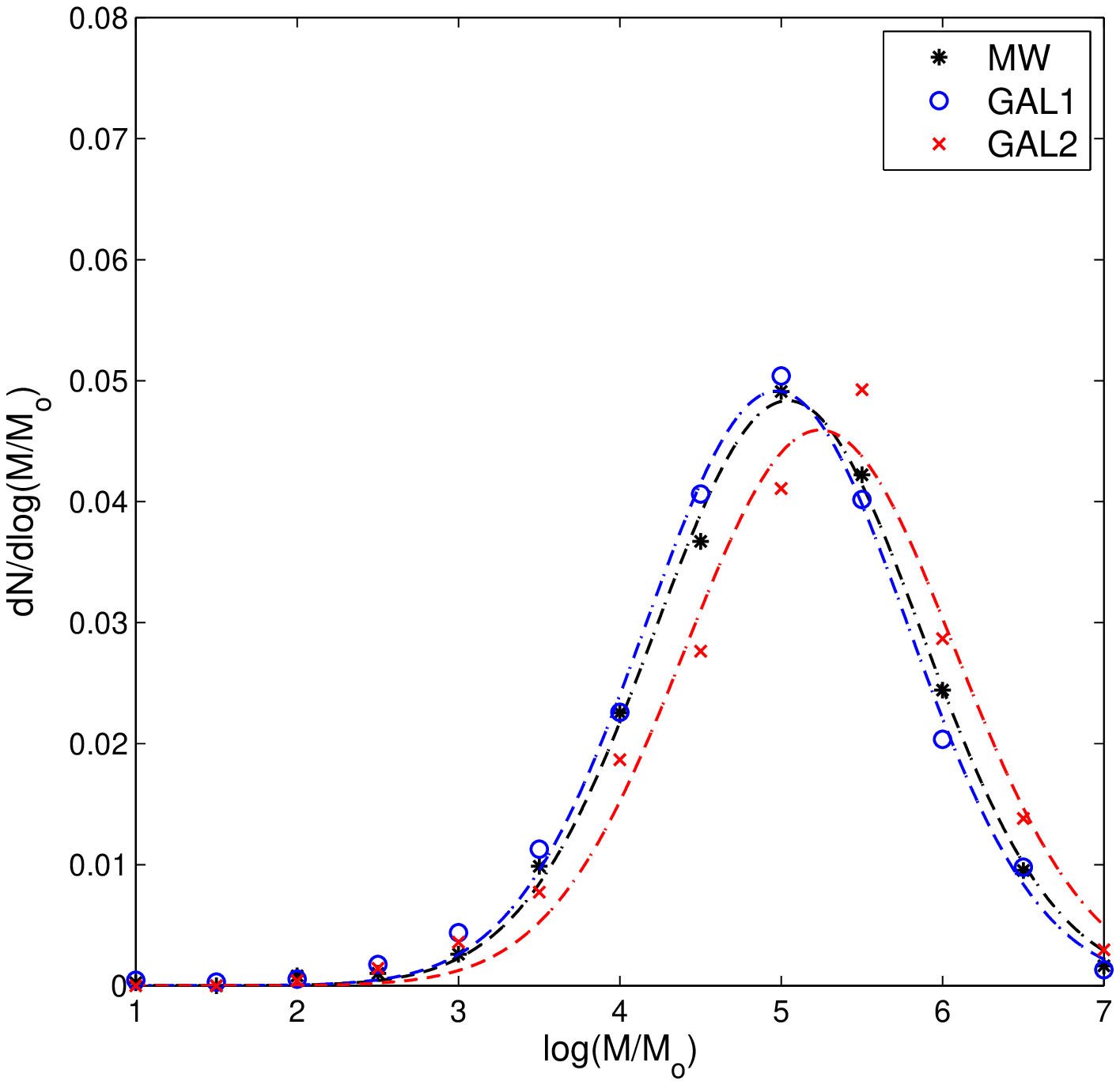}
\caption{Evolution of the star cluster systems in the MW, GAL1 and GAL2 disc galaxy models. \textit{Top panels}: Minimum initial mass of surviving clusters after 13 Gyr of evolution as a function of the galactocenttric distance (left panel) and predicted survival rates as function of the galactocentric distance (right panel). \textit{Bottom panel}: evolved normalized mass function for the  different galaxy models. The points show the results from the simulations, while the lines show the best fit with a log-normal distribution.}
\label{fig:bulge_cluster_systems}
\end{figure}

We included in our database a Milky Way-like galaxy (MW model), an intermediate massive bulge spiral galaxy (GAL1 model) and a massive bulge spiral galaxy (GAL2 model). The mass of the disc and of the dark matter halo is the same for all these models (see Tab. \ref{tab:database}). Fig. \ref{fig:bulge_cluster_systems} shows the properties of the simulated star cluster systems after 13 Gyr of evolution. Firstly, we note that in the inner regions all these models can destroy clusters quite efficiently. In the inner 5 kpc, which roughly corresponds to the radius of the galactic bulge, only clusters more massive than approximately  $10^5$ $M_\odot$ can survive, with the limit increasing from the MW model to the GAL2 model. For increasing galactocentric distances, the minimum survival mass in the MW model decreases faster than in the GAL1 and GAL2 models, reaching the minimum value of the star clusters' mass function at galactocentric distances roughly corresponding to the size of the stellar disc. Accordingly, the survival rate reaches unity at these values as well. Regarding the evolution of the mass function, the distribution of the masses of all the three cases is very well represented by a log-normal function. The MW and GAL1 models mass function are characterized by a smaller value of the turn-over mass, a larger value of amplitude and a smaller value of the dispersion compared to the mass functions in the GAL2 model.
According to these results, we expect the number of stellar clusters in the outer haloes of bulge-disc galaxies to be the same as the initial number, with a survival rate decreasing with decreasing galactocentric distances and reaching less than 20\% in the inner regions dominated by the bulge. 
As a final note, we found that, in spite of the remarkably different mass of the bulge, the MW, GAL1 and GAL2 models affect in an overall similar way the evolution of their hosted cluster system.\\

Regarding the specific case of the MW galaxy model, we found that about 7\% of the initial number of star clusters can survive 13 Gyr of evolution. By comparing this number with the Galactic GC system (counting about 150 objects), it implies that the initial number of  globular clusters is estimated to be about 2150, which, assuming an initial power-law distribution of the masses, translates into a total initial mass of the MW model star cluster system of about $2.9\times 10^8$ $M_\odot$. By comparing the initial mass of the simulated Galactic globular cluster system with the total final mass of the surviving clusters, our results indicate that about $1.3\times 10^8$ $M_\odot$ of the field stellar population was originally formed in star clusters. Assuming that the stellar mass of the galactic halo is $1\%$ of the total stellar mass of the Galaxy \citep{Binney08}, and that approximately $2\%$ of the halo stellar component is contained in surviving globular clusters \citep{Bland-Hawthorn00}, we can estimate the mass of the field stars in the halo of our MW model  to be about $6.9\times 10^8$ $M_\odot$, which translates into the prediction that  $19\%$ of the halo stars originated in globular clusters. Such an estimate is in good agreement with the results of \cite{Martell10}, who concluded that the fraction of halo field stars initially formed within globular clusters may be as large as $50\%$. We also note that \cite{Vesperini10} claimed that the fraction of globular cluster second-generation stars in the Galactic halo would be $<7-9 \%$ for a \cite{Kroupa01} initial mass function, even though it is hard to compare our results with this estimate since we are only considering single stellar population clusters.

\subsection{Star cluster systems in disc galaxies without a massive bulge}
\label{sec:disc_galaxies}
\begin{figure}
\centering
\includegraphics[scale=0.5]{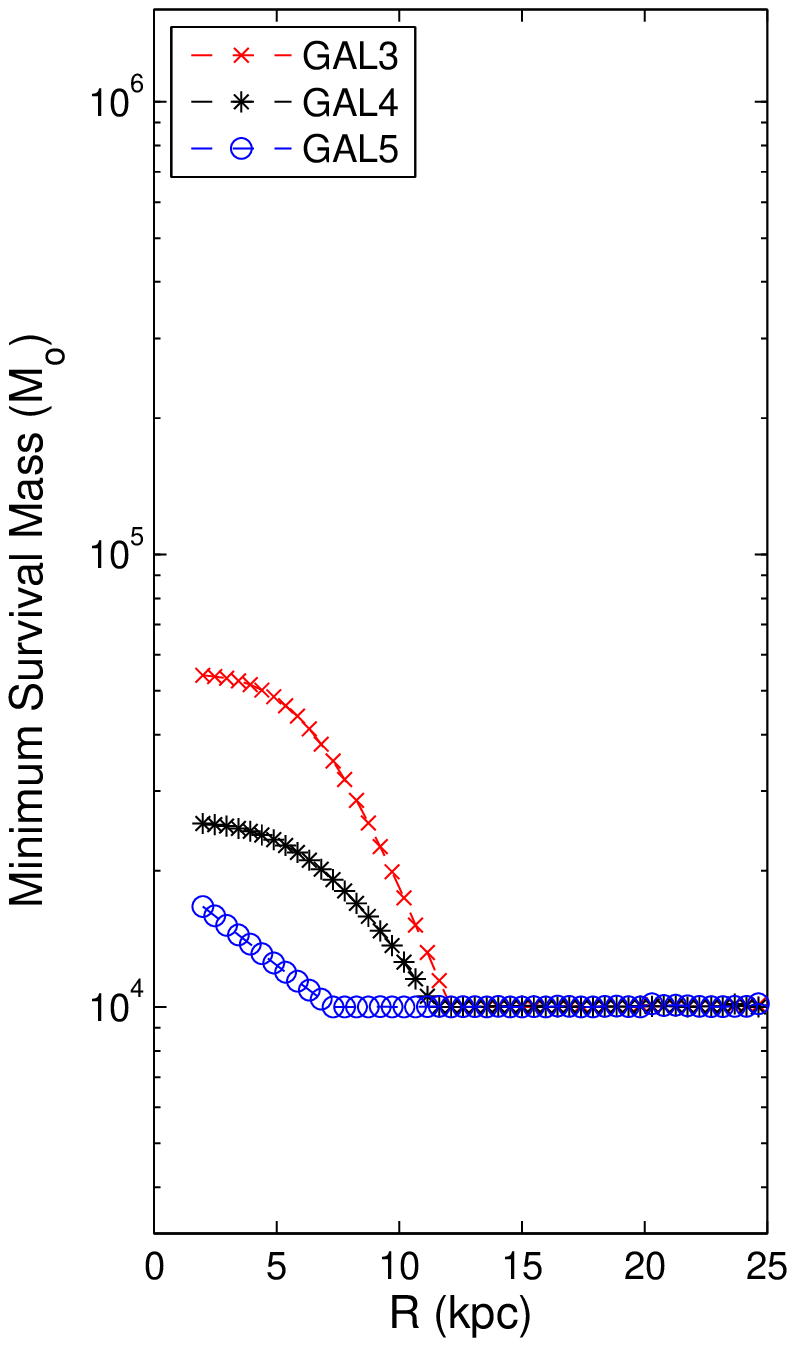}
\includegraphics[scale=0.5]{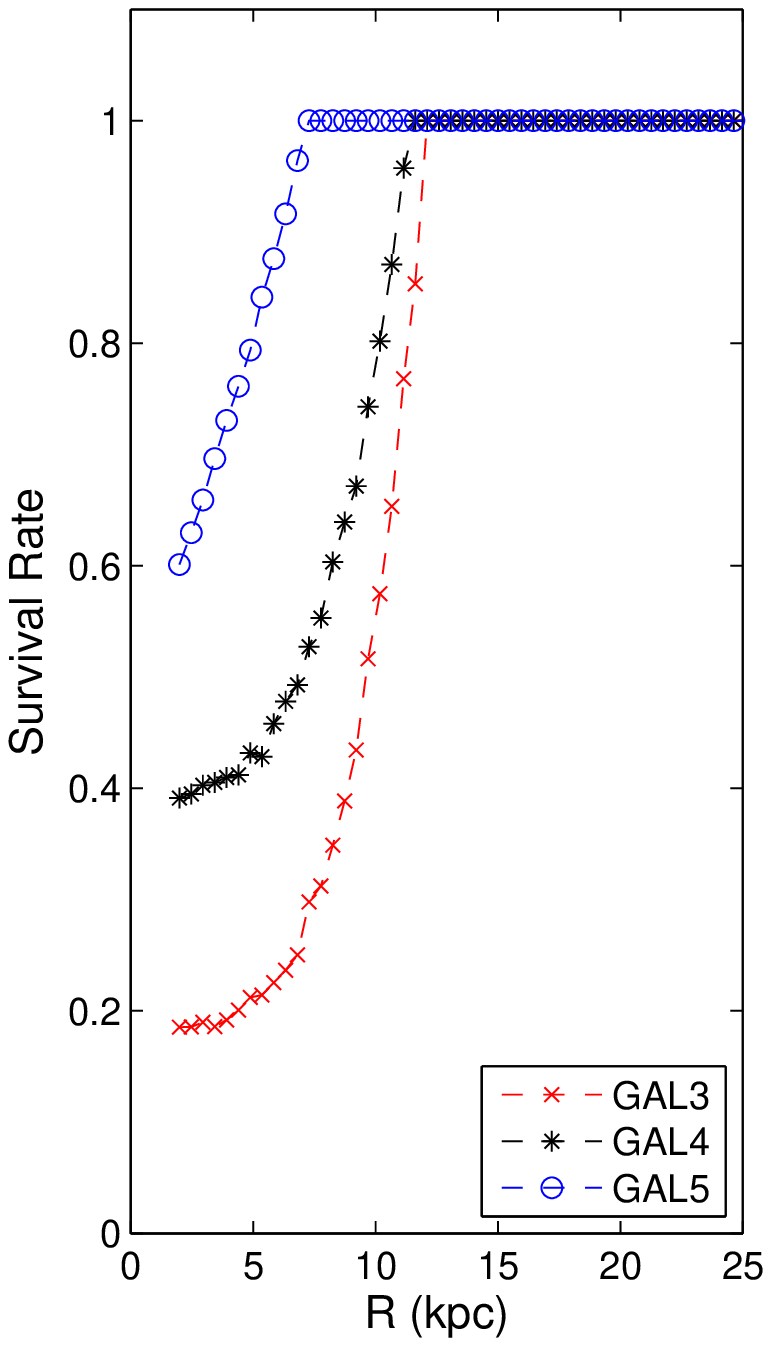}\\
\includegraphics[scale=0.58]{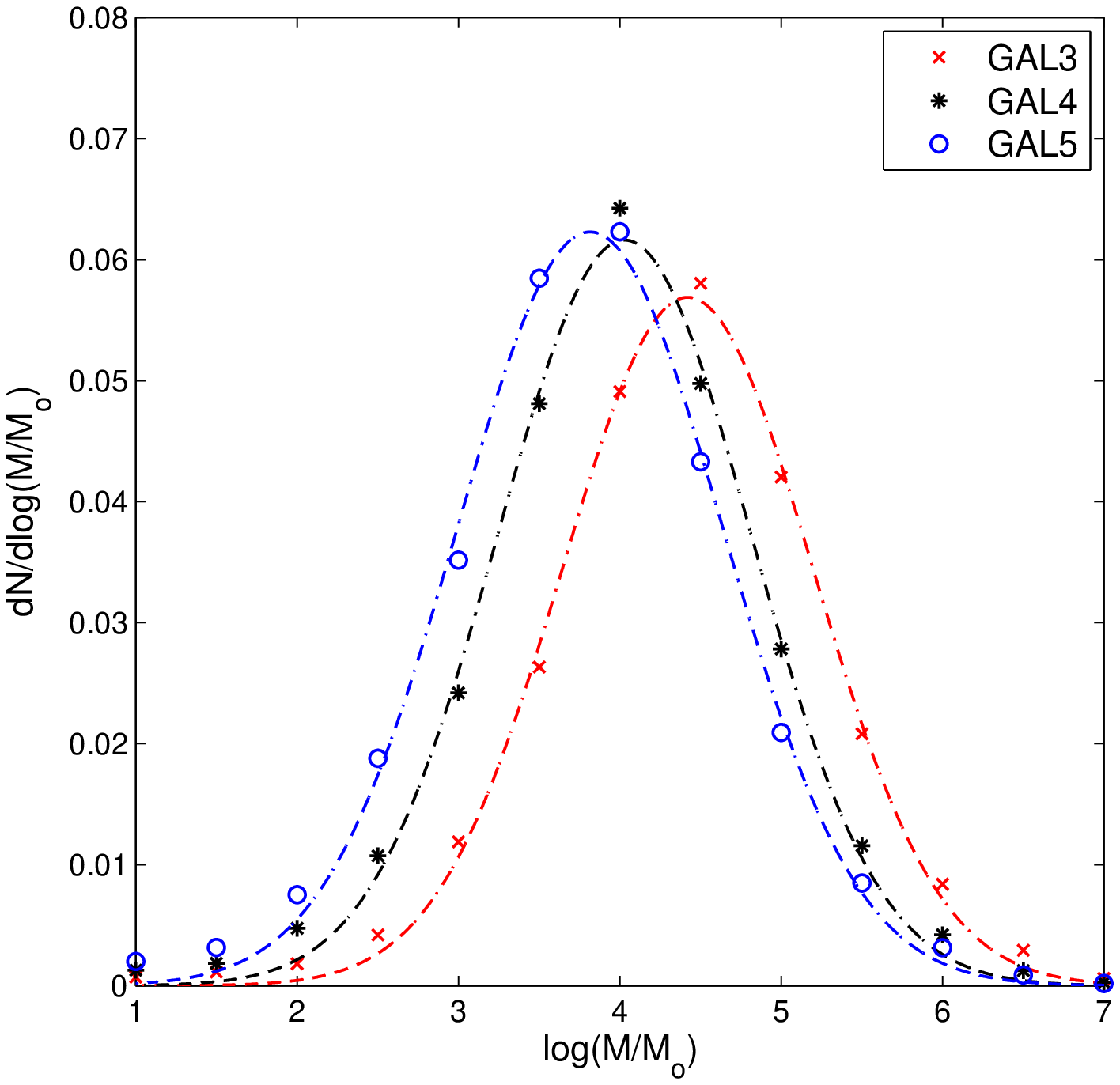}
\caption{Evolution of the star cluster systems in the GAL3, GAL4 and GAL5 disc galaxy models. The content of the various panels is the same as in Fig. \ref{fig:bulge_cluster_systems}.} \label{fig:disc_cluster_systems}
\end{figure}

In the GAL3, GAL4 and GAL5 models we don't include any initial bulge, but only an initially exponential disc with decreasing masses (see Tab. \ref{tab:database}). Firstly, we note that, as evident in Fig. \ref{fig:galaxy_models}, these galaxies experience quite different intrinsic dynamical evolution. In fact, over a short time scale of evolution (typically 0.5 Gyr) all the disc models undergo dynamical instability in the central regions and develop a bar of strength increasing with the mass of the galactic disc. The development of a bar in the inner regions induces an increase of the minimum survival mass of the star clusters in the inner 10--15 kpc of the galaxy models. This result is shown in Fig. \ref{fig:disc_cluster_systems} and is in agreement with the conclusion of Paper II, where we found an enhancement of the mass-loss rate for star clusters evolving in the potential generated by a rotating bar. In fact,  the survival probability is found to increase for decreasing bar strength. For radii greater than approximately 10-15 kpc the minimum survival mass equals the assumed low-mass cut of the initial mass function and the number of star clusters in the evolved system is the same as the initial number. The effect of the different tidal forcing is visible also in the distribution of the masses of the evolved star cluster systems (see Fig. \ref{fig:disc_cluster_systems}). We note that the mass function turn-over mass increases for increasing galactic disc masses, mirroring the higher efficiency in destroying star clusters. By comparing the survival rates obtained for these models with the survival rates obtained for galaxies with massive bulges (Sec. \ref{sec:bulge_galaxies}), it appears that the bulge/bar plays a major role in shaping the properties of the hosted star cluster systems up to galactocentric distances comparable with the radius of the galactic disc, while the halo star cluster systems preserve their initial properties.

\subsection{Star cluster systems in Magellanic-type dwarf galaxies}

\begin{figure}
\centering
\includegraphics[scale=0.5]{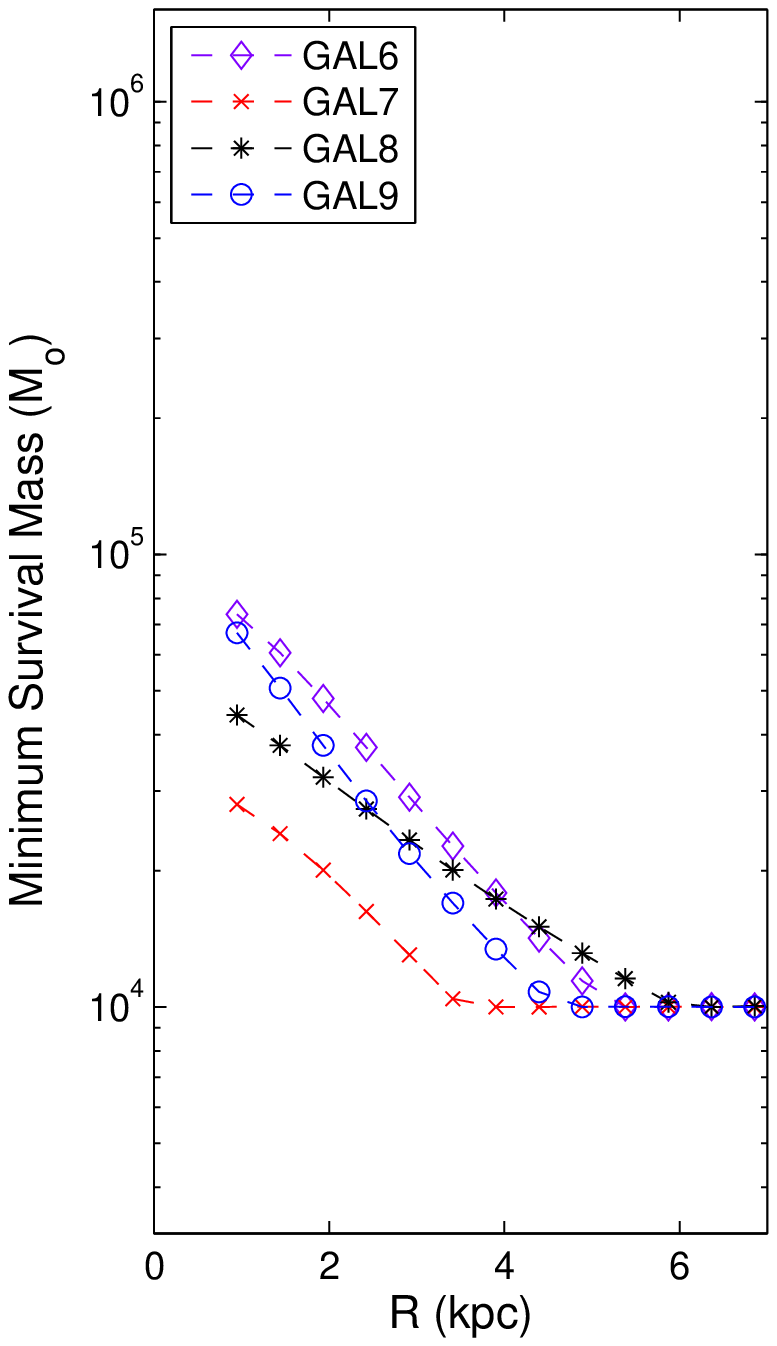}
\includegraphics[scale=0.5]{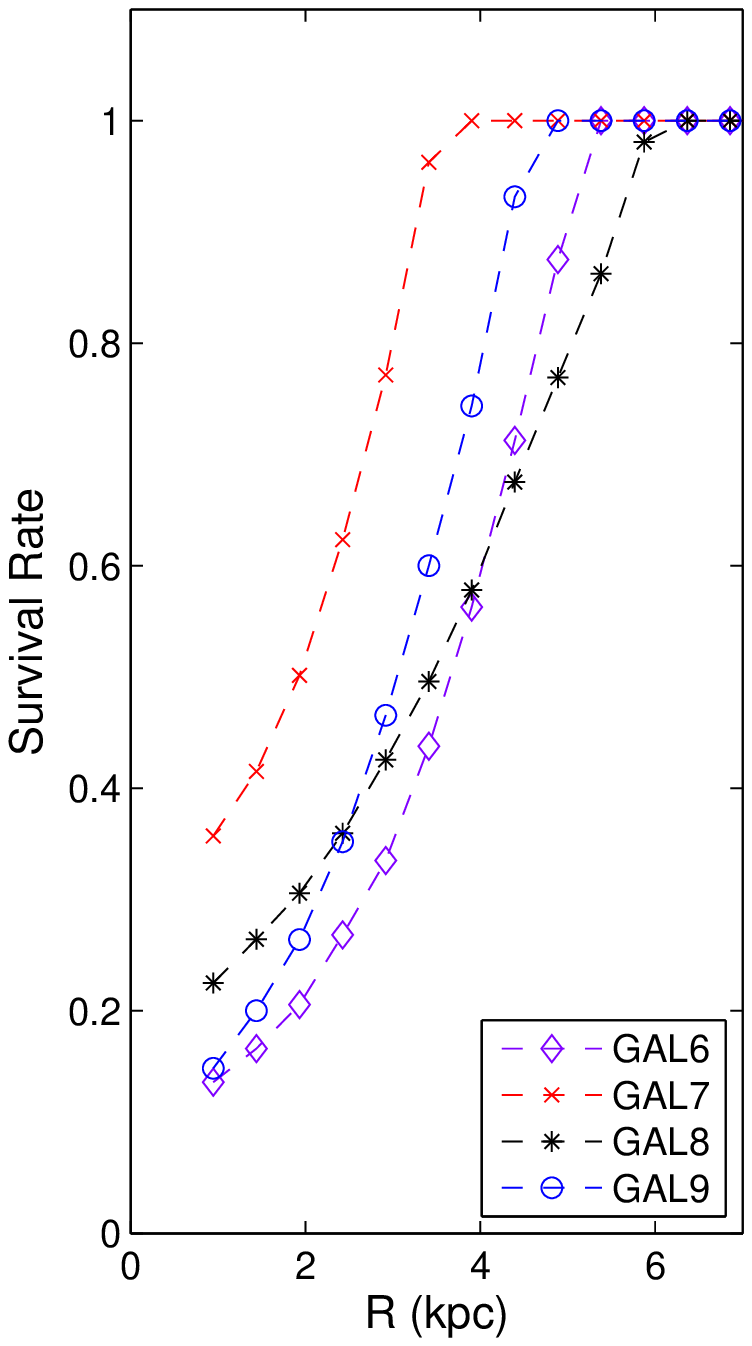}\\
\includegraphics[scale=0.58]{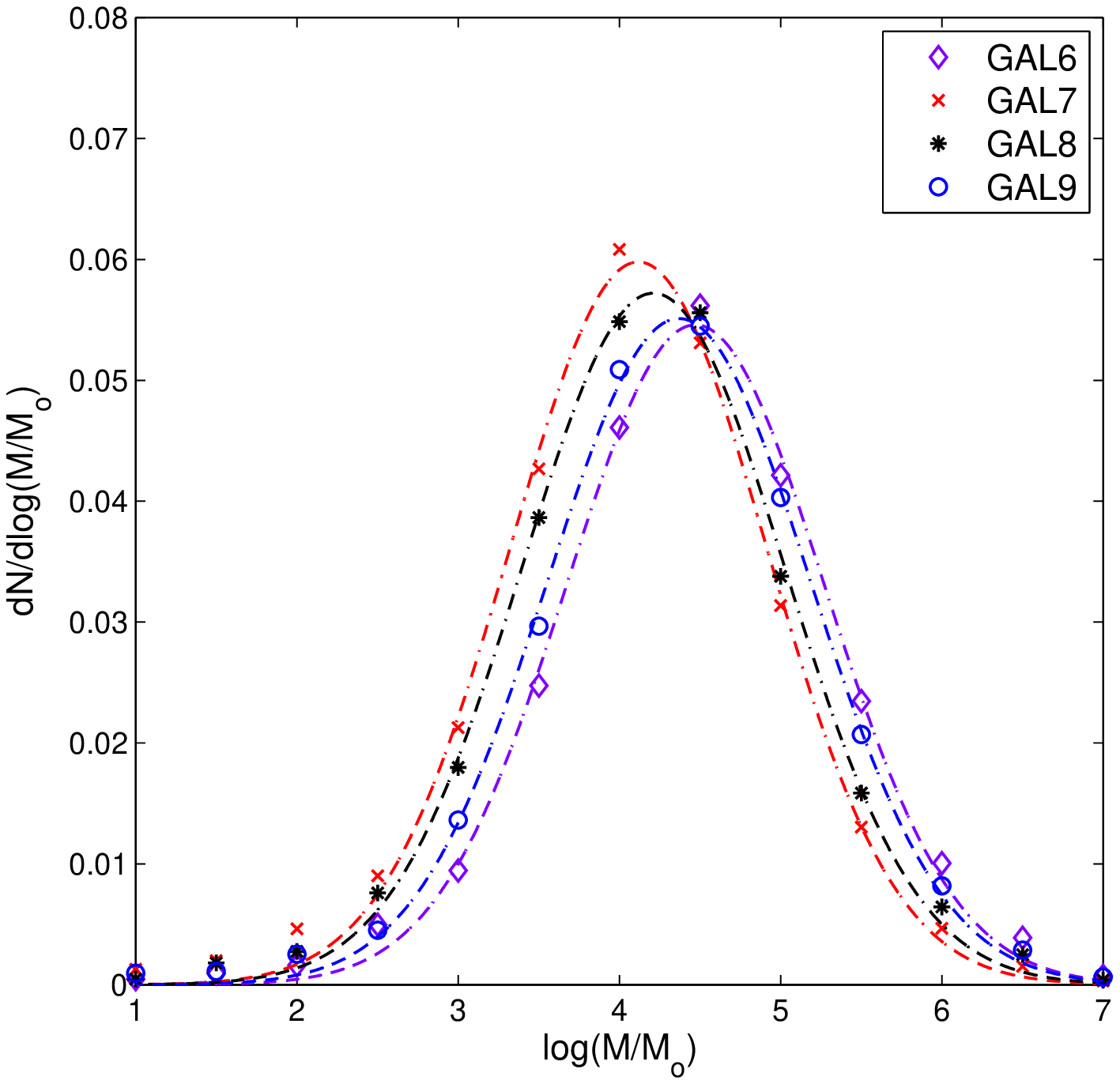}
\caption{Evolution of the star cluster systems in the GAL6, GAL7, GAL8 and GAL9 dwarf galaxy models. The content of the various panels is the same as in Fig. \ref{fig:bulge_cluster_systems}.}\label{fig:dwarf_cluster_systems}
\end{figure}

Our database includes four models of dwarf galaxies (see Tab. \ref{tab:database}). We briefly recall that GAL6 and GAL8 are both models of a LMC disc-like galaxy. The main difference between these models is the mass of the dark matter halo, where GAL6 has a disc to halo mass ratio similar to the one estimated for the LMC, while GAL8 has a lower disc to halo mass ratio. GAL7 is a low-surface-bightness LMC-like dwarf galaxy, with a decreased dark matter halo density/concentration. Finally, GAL9 is a model of a dwarf disc galaxy with disc mass equal to 1/3 of the LMC disc. 
Figure \ref{fig:dwarf_cluster_systems} shows the properties of the simulated star cluster systems after 13 Gyr of evolution. Firstly we note that the results indicate that even dwarf galaxies can quite efficiently destroy star clusters. This effect can be interpreted in light of the results of cosmological simulations, where the dark matter halo of  dwarf galaxies is denser and more centrally concentrated than in disc galaxies. In fact, we found that  models with higher halo mass/concentration are characterised by a higher value of the minimum survival mass, inducing differences in the survival rates. The importance of the galactic mass density is confirmed by the overall similarity of the properties of the evolved star cluster systems of the GAL7 and GAL8 models, which share the same characteristics of the dark halo and differ only for the value of the disc mass. Similarly to what we found for disc galaxies, we expect the initial populations of dwarf galaxies to survive for the whole cosmic time of evolution for galactocentric distances greater than approximately 2-3 times the size of the disc, depending on the mass density of the galactic  environment. Unlike for the case of disc and elliptical galaxies, which have more spatially extended star cluster systems, in the case of dwarf galaxies we are assuming that all the clusters are located within the radius of the stellar component. As our results show, this has a deep impact on the disruption rate of the clusters since the mean density of the environment in which they evolve, which is assumed a priori, is higher.\\
Old globular cluster systems in isolated Magellanic-type dwarf irregular galaxies have been studied extensively by \cite{Georgiev08}. Quite interestingly, the authors found an apparent lack of low-luminosity globular clusters, arguing that such a lack is a consequence of environmental effects, amongst other possibilities. In the same work the authors note that the sizes of old globular clusters in dIrr galaxies are smaller than the size of young globular clusters observed in the Galaxy and in the LMC. Both these forms of evidence are qualitatively consistent with the results of our simulations, which suggest a quite high efficiency for dwarf galaxies in destroying low-mass star clusters due to non-negligible tidal interactions, which could also explain the small sizes of these old objects.

\subsection{Star cluster systems in elliptical E0 galaxies}
\begin{figure}
\centering
\includegraphics[scale=0.5]{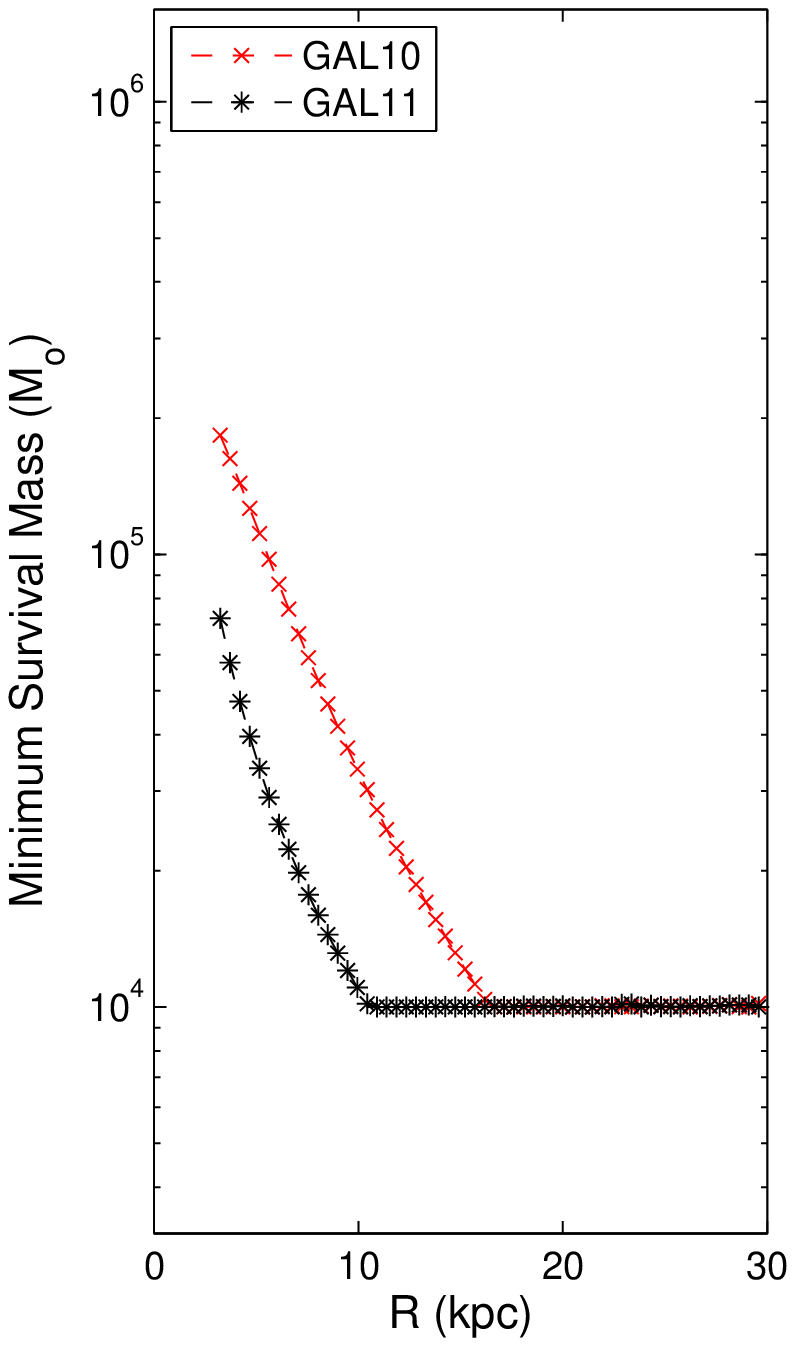}
\includegraphics[scale=0.5]{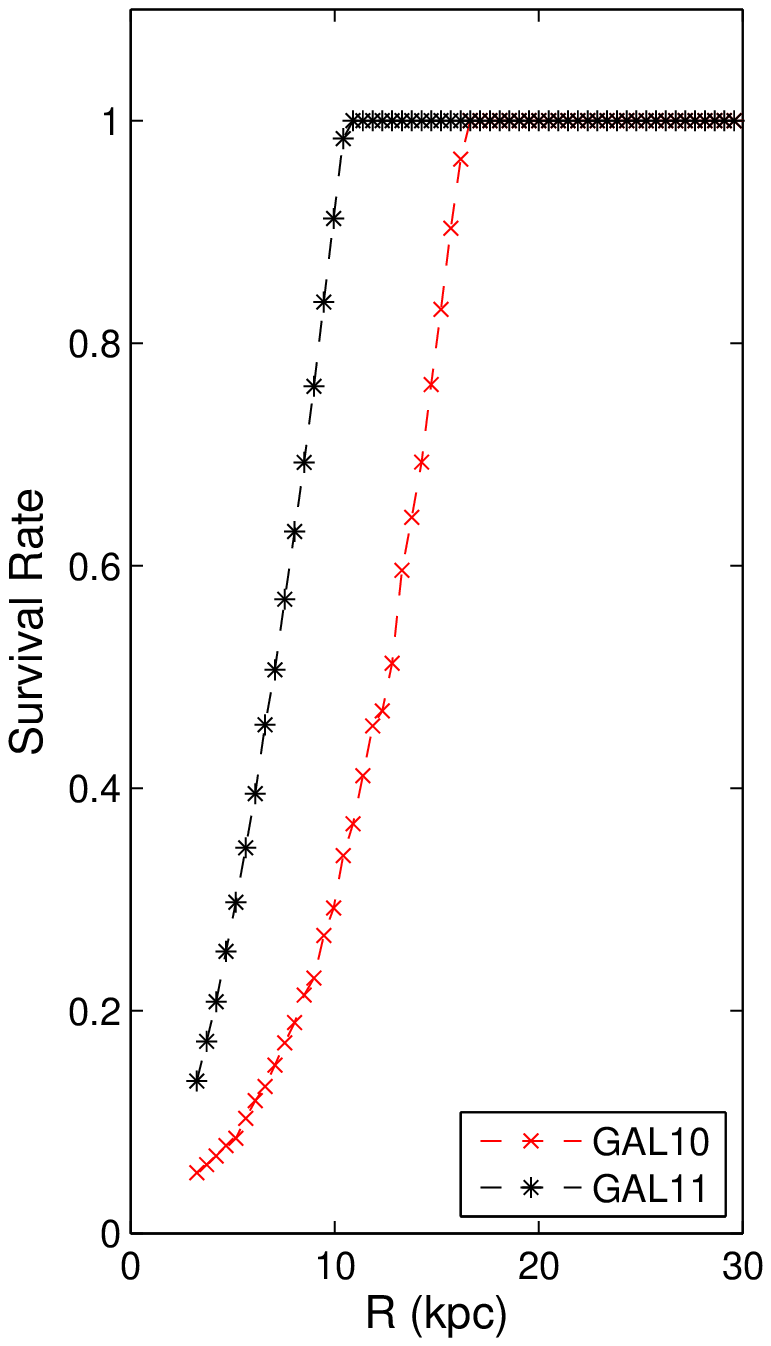}\\
\includegraphics[scale=0.58]{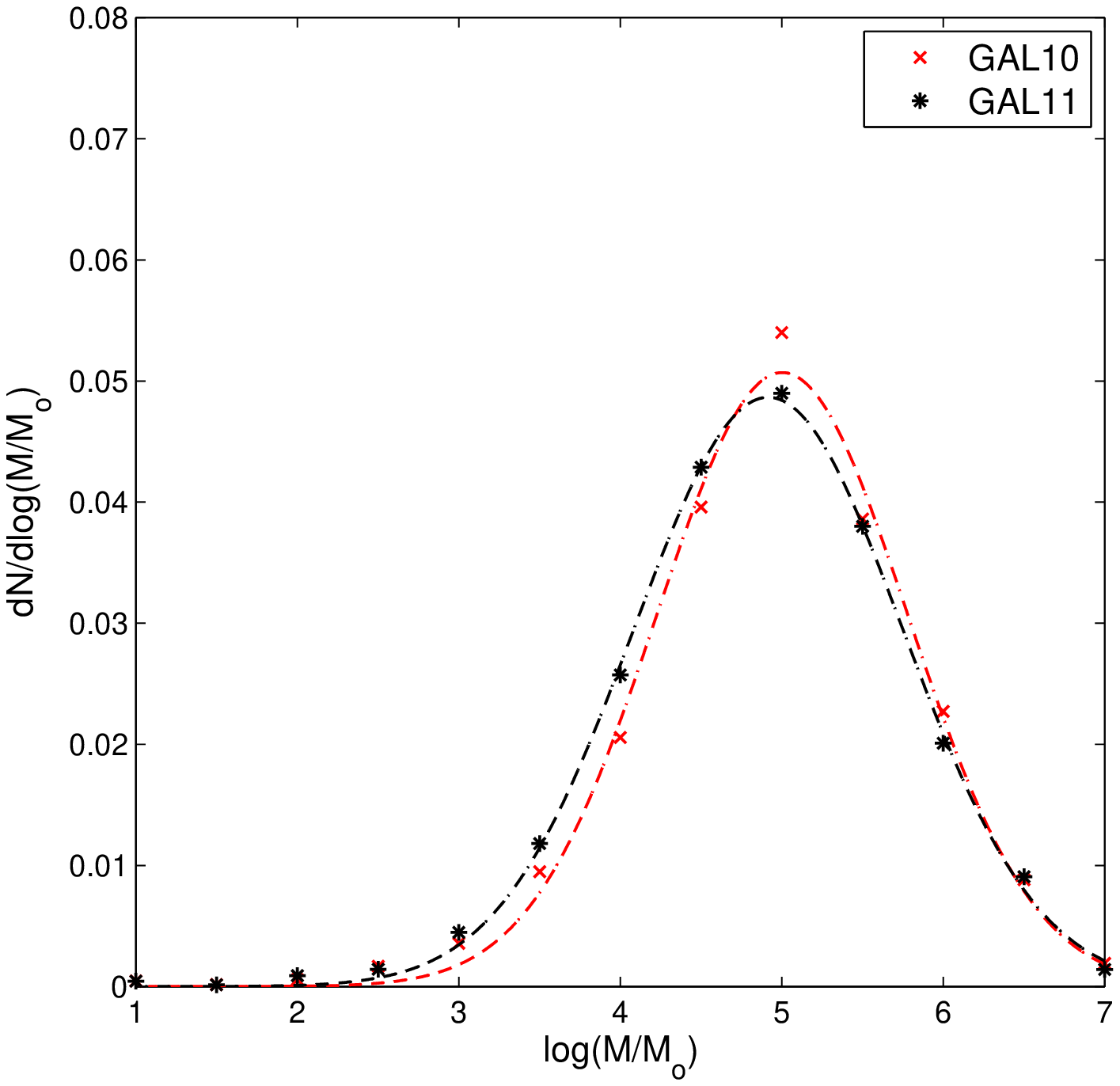}
\caption{Evolution of the star cluster systems in the GAL10 and GAL11 elliptical galaxy models. The content of the various panels is the same as in Fig. \ref{fig:bulge_cluster_systems}.}\label{fig:elliptical_star_cluster_systems}
\end{figure}

We simulated two isolated elliptical E0 galaxies with different masses and sizes (see Tab. \ref{tab:database} for details). When compared with the results for disc and dwarf galaxies, early type galaxies seem to be the most efficient systems in destroying their hosted star cluster systems. Fig. \ref{fig:elliptical_star_cluster_systems} shows the results of our simulations. The minimum survival mass reaches values up to about $10^6$ $M_\odot$ in the inner regions, independently of the mass of the galaxy. In the case of the GAL10 model, which is the most massive, the hosted cluster system is very efficiently destroyed, with less than 40 \% of the original population expected to survive up to distances equal to 2 times the radius of the stellar component of the galaxy. According to our calculations, the initial number of clusters is predicted to survive at distances greater than roughly 4 times the galaxy radius. Also in the case of smaller galaxy masses (GAL11 model) the tidal forcing can efficiently destroy clusters, with the survival rate reaching unity at distances equal to about 4 times the radius of the galaxy. Such a high cluster destruction rate reflects an expected larger value of the evolved log-normal mass function turn-over, increasing with the mass of the galaxy. Our results are qualitative consistent with the conclusions of \cite{Brockamp14}, where the authors studied in detail the erosion of globular cluster systems in elliptical galaxies and demonstrated that early type galaxies can efficiently destroy their hosted star cluster system, with an efficiency dependent on the galaxy's mass, half-mass radius and anisotropy. However,  it should be also noted that star cluster systems in elliptical galaxies could be composed of objects that  formed before and during major merger events, which could have altered the properties of the original populations.

\	\subsection{The choice of the initial size of the star clusters}
\label{sec:initial_size_effect}
The results presented in the previous sections are valid under the assumption that the star clusters form compact ($r_\mathrm{hm} = 1.15$ pc) and with initial sizes independent of the location in the host galaxy. This assumption finds support in the work of \cite{Baumgardt07}, in which the authors concluded that star clusters formed more concentrated than the present day sizes, and expanded by typically a factor 3 or 4 up to a factor 10 as a consequence of gas expulsion activities. However, \cite{Baumgardt10} showed evidence of two distinct populations of star clusters in the Milky Way: compact, tidally under-filling star clusters and tidally filling star clusters. The authors argue that, while tidally under-filling star clusters were born compact with $r_\mathrm{hm} < 1$ pc, the majority of tidally filling star clusters formed with large half-mass radii. In light of this evidence, in the present section we investigate the consequences on the properties of a star cluster population if we instead assume the clusters formed initially filling their tidal radius. We focus on the results for the MW galaxy model as a reference.

\begin{figure}
\centering
\includegraphics[scale=0.5]{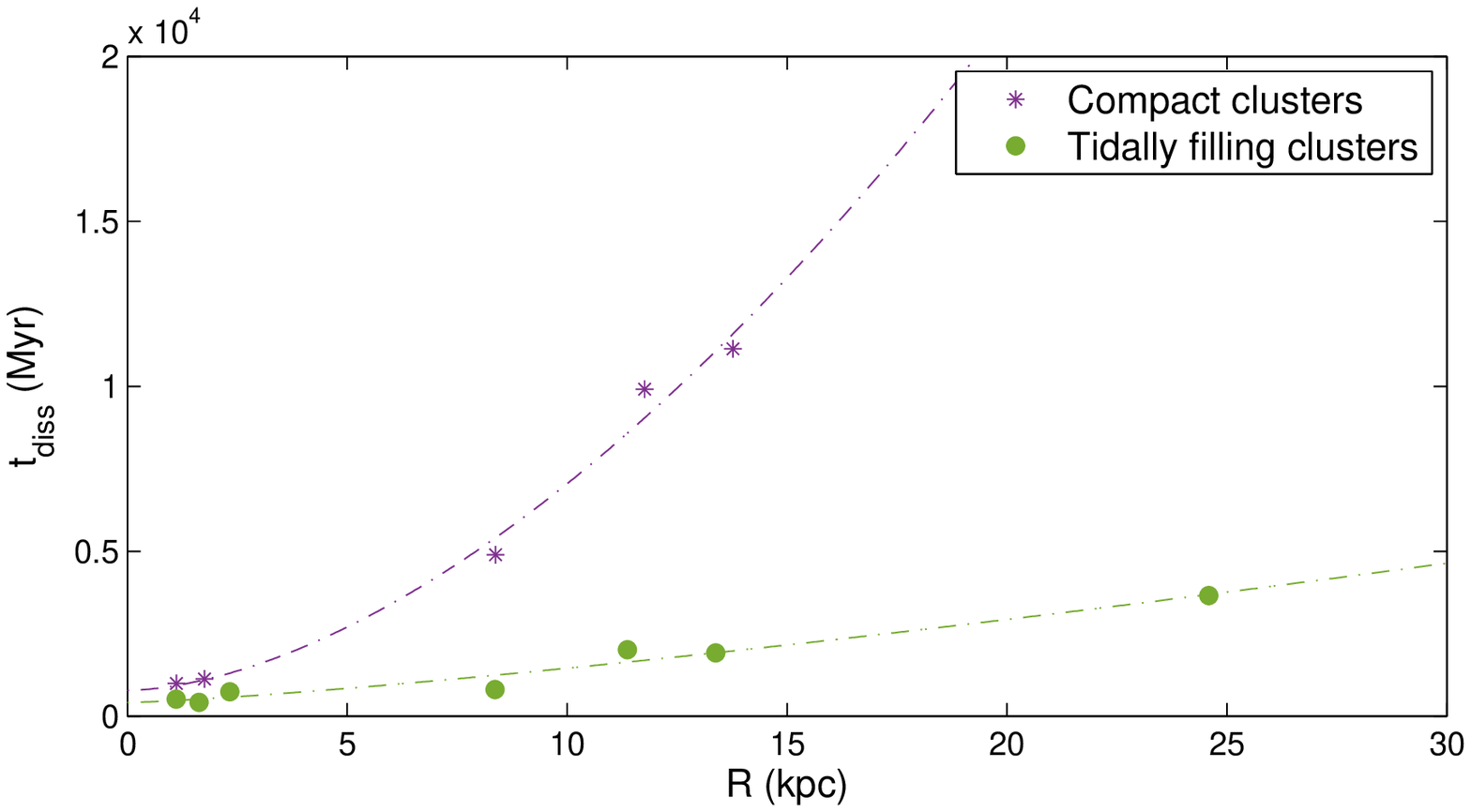}\\
\includegraphics[scale=0.5]{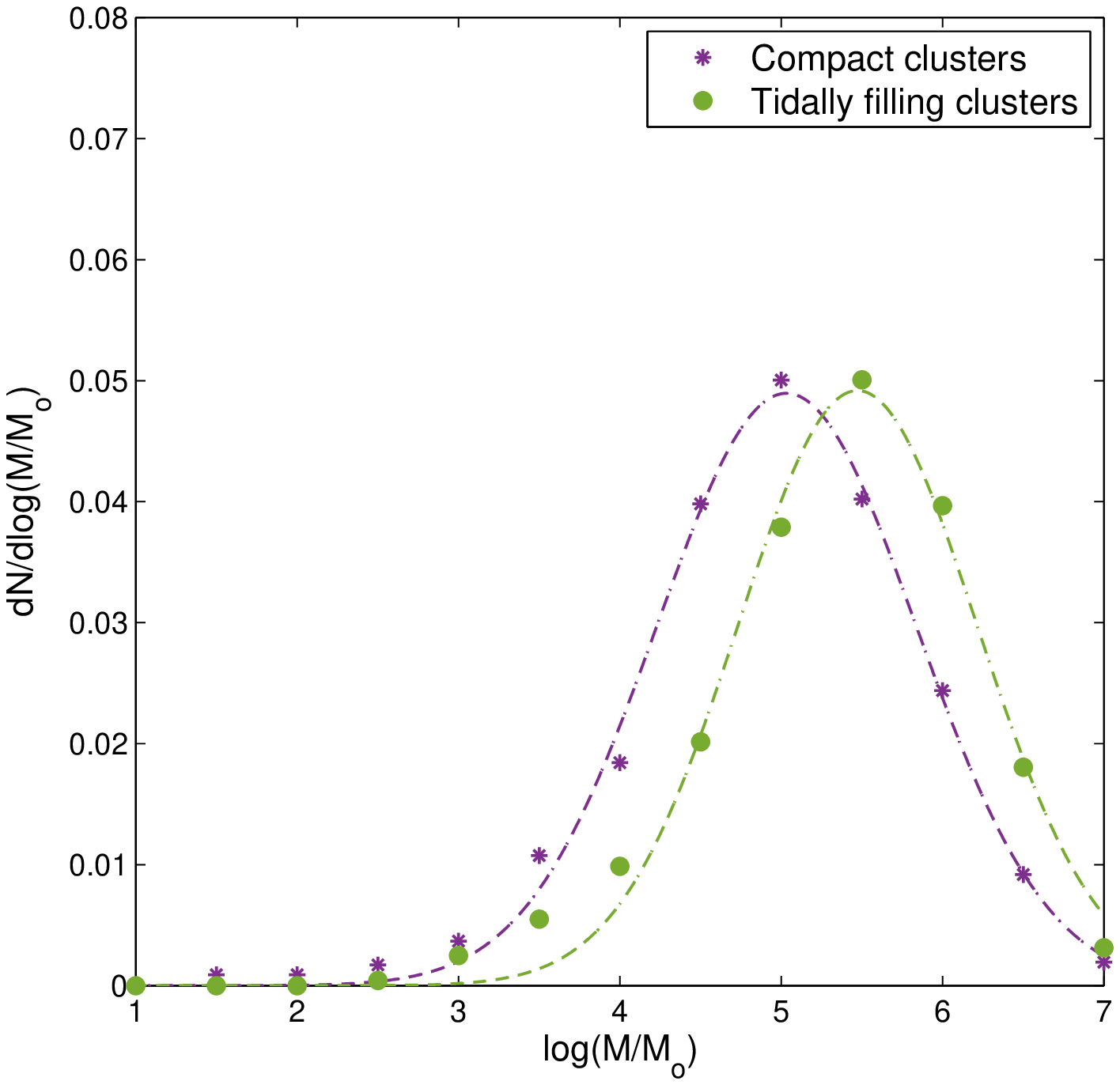}
\caption{\textit{Top panel}: Predicted dissolution times as a function of the orbit-averaged galactocentric distance for initially compact (purple stars) and tidally filling (green filled circles) star clusters with 10k stars evolving in the MW galaxy model. \textit{Bottom panel}: Comparison of the corresponding predicted evolved cluster mass function of the two star cluster systems. The dotted lines represent the fit of the data points with a log-normal distribution.}
\label{fig:initial_sizes}
\end{figure}

We sssumed as a typical ratio between half-mass radius and tidal radius the value of 4 (see Paper I). We then evaluated the instant tidal radius at the start of the simulation for each star cluster as described in Sec. \ref{sec:mesh_size} and set the initial sizes of the clusters (through the choice of the half-mass radius) in order to fill the derived value of the tidal radius. In this case, the sizes of the clusters are of course dependent on their initial position in the galaxy, with sizes increasing for increasing galactocentric distances. The top panel of Fig. \ref{fig:initial_sizes} shows the dissolution times obtained from direct $N$-body simulations for clusters with scaled initial sizes (tidally filling clusters) and for clusters with non-scaled initial sizes (compact clusters). As expected, for each given orbit, initially tidally filling clusters dissolve faster than the initially more compact objects. Such a difference in dissolution times is reflected in different properties of the evolved cluster mass functions (see bottom panel of Fig. \ref{fig:initial_sizes}). As a consequence of the higher destruction rates, the evolved tidally-filling star cluster population is characterised by a higher value of the mass function turn-over ($M_\mathrm{TO}= 2.97\times 10^5$ $M_\odot$) if compared with the mass function turnover of the population of  initially compact clusters ($M_\mathrm{TO}= 1.08\times 10^5$ $M_\odot$).

A further important consideration in the choice of the initial size of the clusters is related to the underling assumption in the derivation of eq. \ref{eq:t_diss}, i.e. the half-mass radius of the star clusters scaling proportional to their tidal radius. Accordingly, ignoring any dependence of the initial size of the clusters on their galactocentric distance could in principle invalidate the use of eq. \ref{eq:t_diss}. However, we found that initially compact clusters rapidly expand within the first 500 Myr to fill their tidal limit, after which the ratio between half-mass radius and tidal radius relaxes to a typical value of 4 for the majority of the evolution of the clusters (see also Paper I). Such evidence validates the use of eq. \ref{eq:t_diss} for initially compact clusters. Furthermore, the prediction of the ratio between the dissolution times from eq. \ref{eq:tdiss_ratio} is found to be in good agreement with the results of $N$-body simulations for both initially compact and tidally filling models. We emphasise that the results in this work are based on initially compact clusters and would need to be re-derived for tidally filling models. For example, the values in Tab. \ref{tab:tdiss_parameter} for the MW galaxy model would be $A=-4.60 \times 10^2$, $B=6.57 \times 10^2$ and $\alpha=0.6$ for initially tidally filling clusters.

Finally, according to the results of \cite{Baumgardt10}, we note that a more realistic cluster population would be composed of both initially compact and more extended objects, with properties expected to be somehow intermediate between the two extreme cases presented in this section.

\section{Cluster mass function and galactic environment}
\label{sec:discussion}
One of the main results obtained is that an initial cluster power-law mass function evolves into a mass distribution which is extremely well represented by a log-normal function, independently of the galactic environment. In this sense, our results are in agreement with the conclusion that an initial power-law cluster mass function evolves into an ``approximately'' universal present-day mass function. However, as intuitively expected, different galaxies are found to be able to destroy star clusters with different efficiencies. Assuming that all the SCSs are characterized by the same universal initial conditions, a good indicator of the star cluster destruction efficiency of the host galaxy is represented by the value of the turn-over of the log-normal mass function \citep[e.g.][]{Vesperini97,Vesperini98}. According to results from \cite{Jordan06} and \cite{Harris14}, not only the turn-over mass, but also the dispersion of the masses correlates with the properties of the host galaxy, in the sense that SCSs in smaller galaxies have narrower luminosity functions (at least for the case of early-type galaxies.) 

\begin{figure}
\hskip -5mm
\includegraphics[scale=0.5]{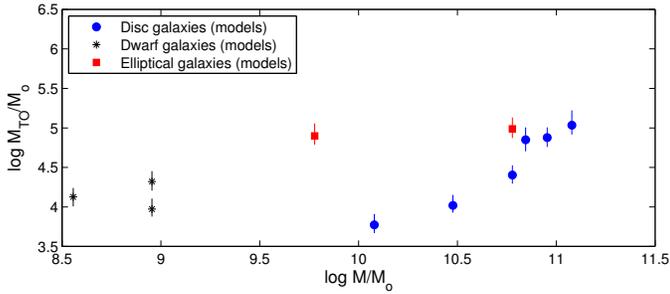}
\caption{Scatter plot of the distribution of the predicted turn-over masses of the evolved star cluster systems as a function of the stellar mass of the host galaxy. The blue dots, black stars and red squares show the results for disc galaxies, Magellanic-type dwarf galaxies and elliptical galaxies, respectively.}
\label{fig:TO_mass}
\end{figure}

\begin{figure}
\hskip -10mm
\begin{tabular}{r}
\includegraphics[scale=0.53]{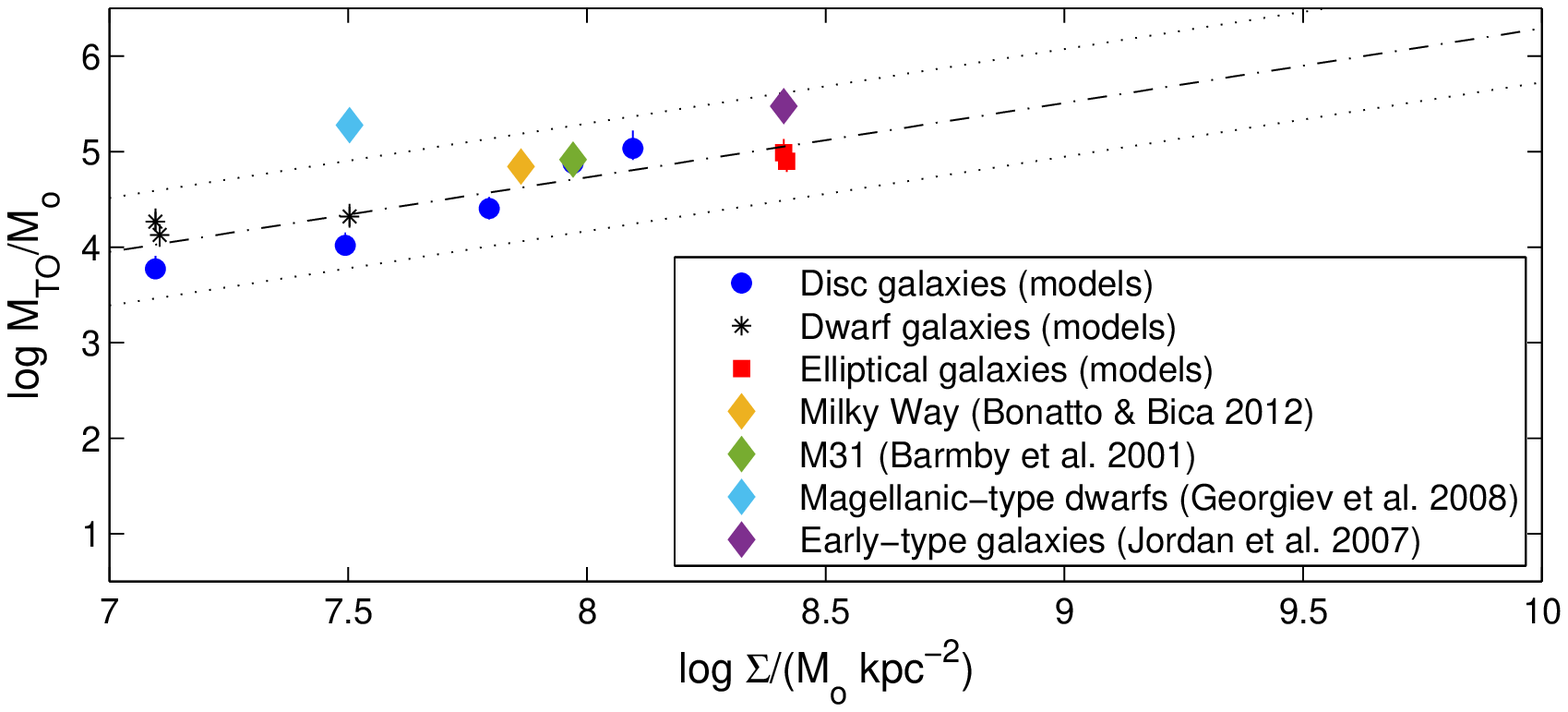}\\
\includegraphics[scale=0.53]{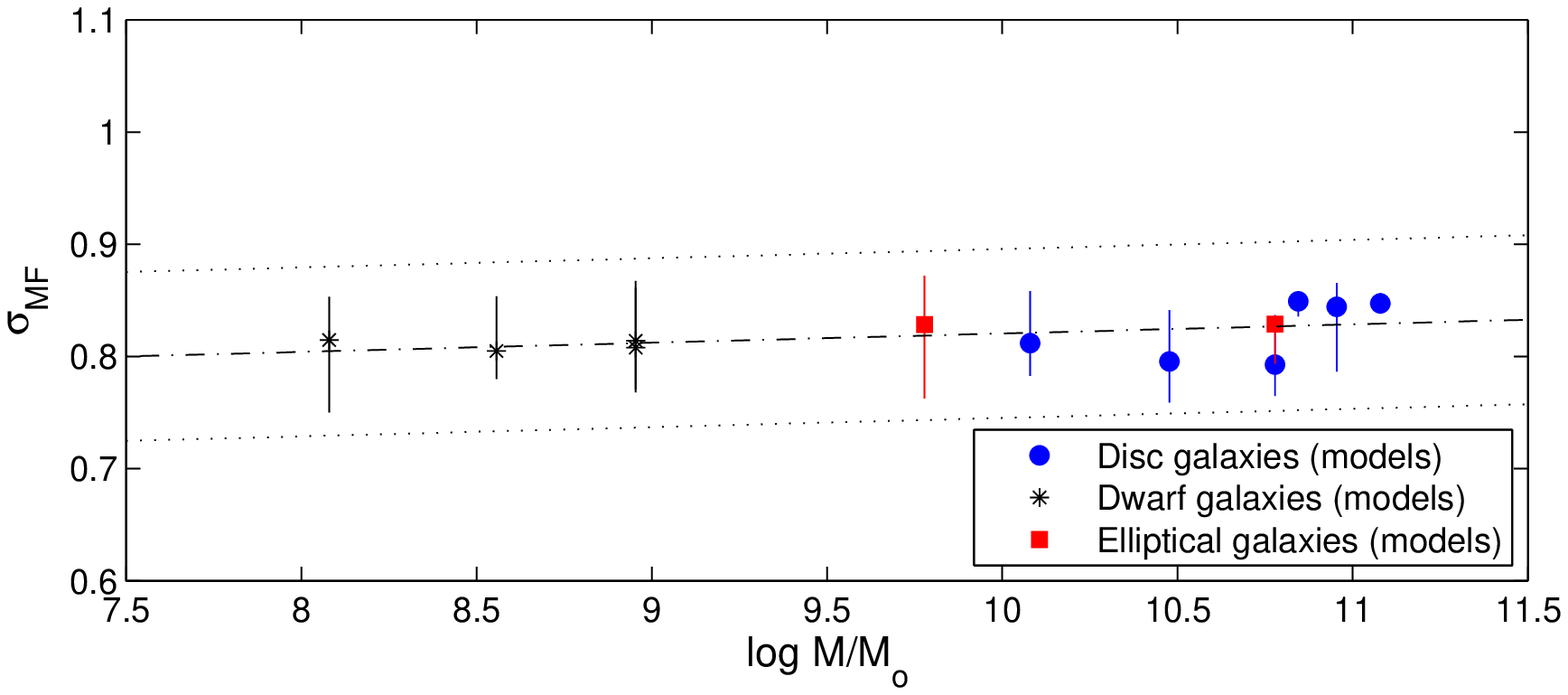}
\end{tabular}
\caption{\textit{Top panel}: Evolved cluster mass function turn-over as a function of the mean surface stellar mass density of the host galaxy for disc galaxies (blue dots), Magellanic-type dwarf galaxies (black stars) and elliptical galaxies (red squares). The diamonds represent the data point from observations (see the legend for the references). The black dot-dashed lines shows the result of a linear regression model fit, while the black dotted lines show the $3\sigma$ errors associated to the fit. \textit{Bottom panel}: Correlation between dispersion of the mass function and stellar mass of the host galaxy. The meaning of the symbols is the same as in the top panel. The dot-dashed black line and the black dotted lines show the best linear regression model and the associated $3\sigma$ errors, respectively.} 
\label{fig:MF_galaxy_relation}
\end{figure}

Our results do not highlight any strong correlation between the properties of the evolved cluster mass function and the stellar mass or the morphological type of the host galaxy (Fig. \ref{fig:TO_mass}), although we could argue that there is a correlation for the disc galaxies. 
Instead, the cluster mass function turn-over correlates well with the average stellar mass surface density $\Sigma$ of the host galaxy, as visible in the top panel of Fig. \ref{fig:MF_galaxy_relation}. A fit with a linear regression model produces the empirical relation
\begin{equation}
\log_{10}(M_\mathrm{TO}) = a + b \log_{10}(\Sigma)\;\;\;,
\end{equation}
where $a = -1.59$ and $b = 0.79$. The simulations show that SCSs evolving in denser galactic environments, such as elliptical galaxies, are expected to be more efficiently eroded than SCSs living in disc and Magellanic-type dwarf galaxies. This dependence is in agreement with \cite{Ostriker97}, stating that ``the amplitude of the tidal shocks scales with the density of matter in the galaxy at the position of globular clusters''. 

On the other hand, the dispersion of the mass function is found to weakly correlate with the galactic stellar mass $M_\mathrm{gal}$, roughly following a trend of the form
\begin{equation}
\log_{10}(\sigma_\mathrm{MF}) = c + d \log_{10}(M_\mathrm{gal}/M_\odot)\;\;\;,
\end{equation}
where $c=0.77$ and $d = 0.004$. 
Higher turn-over masses, and hence higher environment densities, are expected to be associated with higher values of the dispersion of the clusters' mass distribution.
The uncertainties associated with the turn-over mass and the dispersion of the mass function were obtained by propagating the uncertainty in the slope of the initial cluster mass function, which was then assumed to be $\gamma = 2.0 \pm 0.2$.
Finally, we note that these results are overall  in good agreement with the observations for SCSs in different galactic environments. As a comparison, we considered the turn-over of the mass function of globular clusters observed in the Milky Way \citep{Bonatto12}, in M31 \citep{Barmby01}, in Magellanic-type dwarfs \citep{Georgiev08} and in early-type galaxies \citep{Jordan06}. The value of the turn-over mass in the case of M31 and of the Magellanic-type dwarfs was obtained by applying the power-law mass-to-light ratio from \cite{Bonatto12} to the turn-over magnitude from \cite{Barmby01} and \cite{Georgiev08}, respectively. In this case we  assumed the GAL1 model with a massive barred bulge to represented to first approximation a model of M31. The predicted value of the mass function turn-over for Magellanic-type dwarfs does not agree as well with the observations. However, Magellanic-type dwarfs are interacting systems in which the interplay between tidal destruction and the ongoing formation of more massive clusters could contribute to explain the observed inconsistency. We also note that the choice of the mass-to-light ratio has a fundamental impact when converting the observed  magnitudes-luminosity into mass.

\begin{figure}
\hskip -8mm
\begin{tabular}{r}
\includegraphics[scale=0.52]{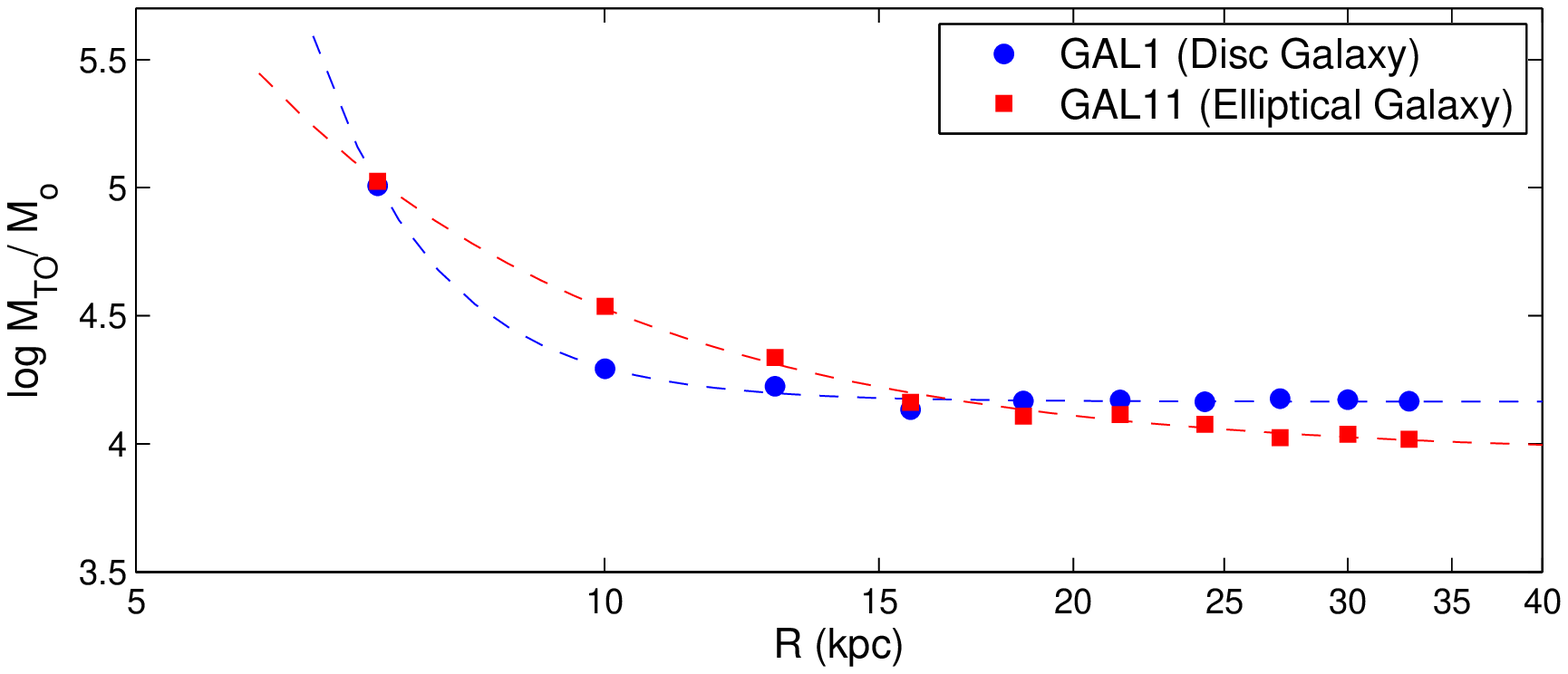}\\
\includegraphics[scale=0.52]{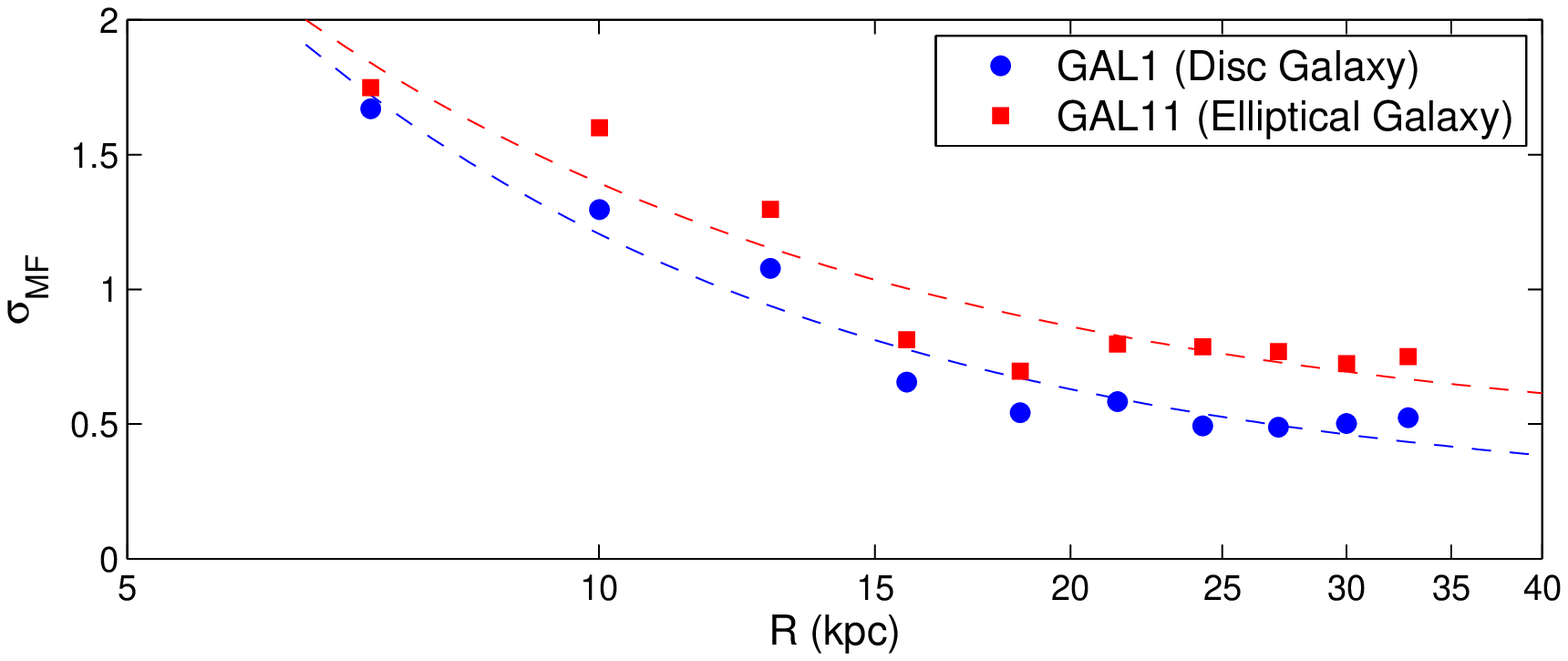}
\end{tabular}
\caption{Radial dependence of the mass function turn-over of the star cluster population for the GAL1 model (blue circles) and the GAL11 model (red squares). The dotted lines show the fit with a power-law function.}
\label{fig:MTO_vs_R}
\end{figure}

Our approach also allows us to investigate the radial dependence  of the mass function in different environments. \cite{Harris14}, for example, found a decrease of the turn-over luminosity $L_0$ for increasing galactocentric distances in the inner regions of their sample of elliptical galaxies, with a weak dependence  $L_0 \propto R^{-0.2}$  in the outer halo. In the present study we evaluated the radial dependence of the mass function turn-over for a disc galaxy (GAL1 model) and for an elliptical galaxy (GAL11 model). The results are shown in Fig. \ref{fig:MTO_vs_R}. In good agreement with \cite{Harris14}, we found that the mass function turn-over is expected to decrease with the galactocentric distance in the inner regions of the galaxy, with a weaker dependence for decreasing distances $R$. We also note that in our models the mass function is dominated by clusters located towards the inner regions of the galaxies, where the number density is higher. On the contrary, the models predict a dependence of the mass function dispersion on the galactocentric distance, which is not observed. Finally, a power-law function is a suitable representation of the data points for both the mass function turn-over and dispersion.

\section{Discussion and conclusions}
\label{sec:summary}
We have developed an alternative method to simulate the dynamical evolution of star clusters in arbitrary tidal fields by coupling collisionless galaxy-scale $N$-body simulations with fully-collisional $N$-body simulations of star clusters, the latter performed with \textsc{nbody6}. The main improvement of our method with respect to other approaches is that we overcame the approximation of linearised forces by reconstructing the gravitational field around the star clusters by applying a 3D polynomial fit procedure. The method is in principle applicable to arbitrary gravitational fields generated by galaxies with different morphologies and in different evolutionary stages, such as mergers and tidal interactions. 

This method was then applied to estimate the impact of different realistic galactic environments on the dynamical evolution of the hosted SCSs. We generated a database of galactic models including disc galaxies, Magellanic-type dwarf galaxies and elliptical galaxies. For each galaxy model, we calibrated generalized evolutionary equations predicting the dissolution time and mass evolution of star clusters as functions of their orbit-averaged galactocentric distance.  As a next step, for each galaxy model we generated synthetic star cluster systems for which the initial properties are assumed to be independent of the galactic environment in which they have formed and, most importantly, share the same universal initial mass function, assumed to follow a power-law distribution with index $\gamma = - 2 \pm 0.2$. By applying the evolutionary model, we have been able to predict survival rates and minimum survival masses of the evolves SCSs as a function of the clusters' orbit-averaged galactocentric distance for the different galaxy models in the database. We confirmed that, independently of the galaxy type, star clusters are preferentially destroyed in the inner region of the galaxies and are more likely to survive in the galaxies' haloes. However, the intensity of the erosion of the star cluster systems appears to be strongly dependent on the properties of the host galaxy. Disc galaxies can destroy star clusters up to galacticentric distances comparable with the size of the galactic disc, with survival rates decreasing as the stellar mass of the galaxy increases. Furthermore, the development of time-dependent components, such as a bar, proved to influence the survival rate of the star clusters. The results showed that, in spite of their lower stellar mass component, Magellanic-type dwarf galaxies can destroy the hosted SCS quite efficiently, with the concentration and mass of the dark matter halo influencing the survival rates of star clusters. Finally, elliptical galaxies proved to be the most efficient systems in destroying clusters.

We investigated the impact of the initial size of star clusters on the prediction of the evolutionary model by comparing a cluster system in which the dependence of the initial size on the galactocentric distance is neglected with a cluster system in which the initial sizes are chosen to fill the tidal radius of the clusters, with both system evolving in the MW galaxy model. The results suggest that the initial size of the clusters has some influence on the evolution of the star cluster systems, although this effect can be considered negligible if compared with the impact of varying the external tidal field.
  
The differing impact of the properties of the external galaxy on the evolution of the star clusters translates into different expected properties of the evolved cluster mass functions. As a first result, we found that in every galactic model an initial power-law mass function evolves into a mass distribution which is extremely well represented by a log-normal function. In this sense, a log-normal mass function is a ``universal'' feature of evolved star cluster systems in galaxies. However, the properties of the log-normal distribution, such as turn-over mass and mass dispersion are proven to be dependent on the properties of the host galaxy. The results of our analysis didn't highlight any strong correlation between the turn-over mass of the cluster mass function and the mass of the host galaxy. On the other hand, we found a dependency of the turn-over mass on the mean density of the stellar mass component of the host galaxy, with increasing $M_\mathrm{TO}$ for increasing environment densities. Furthermore, the dispersion of the mass function is expected to correlate with the stellar mass of the host galaxy, resulting in higher dispersions for increasing masses. An environmental dependence of the mass function has been observed in early-type galaxies \citep{Jordan06,Harris14}. In both these works the authors found an increase of the mass-function turn-over and dispersion for increasing galaxy masses, which is qualitatively consistent with our prediction. On the side, we also note that \cite{Nantais06} found that the luminosity function of NGC 300, which is  morphologically similar to our GAL5 model, may have an intrinsically fainter luminosity function turn-over than the Milky Way or M31, in good agreement with our results. At first glance, the results of the simulations seem to be in good agreement with the observations. A more systematic study investigating the properties of the cluster mass function in different galaxies along the Hubble sequence would be required in order to draw more conclusive results.

The present analysis has however faced several limitations. Firstly, in the simulations we assume the clusters to be coeval and in initial virial equilibrium, neglecting any physical mechanisms leading to their formation. Furthermore, the structure of the galaxies is set at the start of the simulations, implying that we didn't model the galaxies' formation processes, which would require the use of cosmological-scale $N$-body simulations. Also, in the present form the derived evolutionary equations allow us to predict to good accuracy only the dissolution time and the mass evolution of the star clusters, neglecting the evolution of structural parameters and dynamical effects such as half-mass radius and core collapse for which direct $N$-body simulations are still required. We also note that, in order to gauge the dependence of the evolved cluster population properties on the galactic environment, we assumed that the initial properties of the simulated star cluster systems, such as the radial number density distribution, are identical in every galaxy. However, the radial density distribution, which according to our results is expected to strongly influence the subsequent evolution of the whole cluster population, may not be universal within each group of galaxies. Finally, a more systematic comparison of predictions and observations of the evolved mass function properties of extragalactic star cluster systems in different galactic environment is required. 

The present method to couple $N$-body simulations of galaxies and stellar clusters is in principle adaptable to a range of physical scenarios. Considerations should be given to the size of the mesh, the number of grid points and even the possibility to increase the degree of the interpolating 3D polynomia, based on the initial masses and sizes of the clusters of interest and/or in cases of rapidly evolving potentials, such as the ones associated with galaxy mergers and tidal interactions.

As a main conclusion, the present work showed that star cluster systems sharing the same initial properties, such as radial number density distribution and initial mass function, are expected to evolve in different ways in host galaxies with different properties. Under the assumption of a universal power-law initial mass function, the distribution of the masses of the evolved star cluster system is extremely well described by a log-normal function, independently of the external tidal field. However, the intrinsic properties of the log-normal distribution, such as turn-over mass and dispersion, are expected to be dependent on the properties of the host galaxy.

\section{Acknowledgements}
We thank the anonymous referee for comments which we believe have contributed to an improved submission. We also would like to thank Chris Power for productive discussions. This work was performed on the swinSTAR and gSTAR supercomputers at Swinburne University of Technology funded by Swinburne and the Australian Government's Education Investment Fund. LR acknowledges a CRS scholarship from Swinburne University of Technology.

\bibliography{biblio}
\bibliographystyle{mn2e}

\label{lastpage}
\end{document}